\documentclass[aps,prd,11pt,twoside,tightenlines,nofootinbib,showpacs,preprint,superscriptaddress]{revtex4-1}
\usepackage[colorlinks=true]{hyperref}
\usepackage{xcolor}
\hypersetup{colorlinks=true, citecolor=blue, urlcolor=blue, linkcolor=blue}
\usepackage{amsmath,amssymb}
\usepackage[final]{graphicx}
\usepackage{subcaption} 
\usepackage{float}

\begin{document}
\title{Role of $a_0(980)$ in the decays $D^{0} \rightarrow K^{+} K^{-} \eta$ and $\pi^{+} \pi^{-} \eta$}

\author{Sara Rahmani}
\affiliation{School of Physics and Electronics, Hunan University, 410082 Changsha, China}

\author{Wei Liang}
\email{212201014@csu.edu.cn}
\affiliation{School of Physics, Hunan Key Laboratory of Nanophotonics and Devices, Central South University, Changsha 410083, China}

\author{Yu-Wen Peng}
\affiliation{School of Physics, Hunan Key Laboratory of Nanophotonics and Devices, Central South University, Changsha 410083, China}

\author{Yu Lu}
\affiliation{School of Physics, Hunan Key Laboratory of Nanophotonics and Devices, Central South University, Changsha 410083, China}

\author{De-Liang Yao}
\affiliation{School of Physics and Electronics, Hunan University, 410082 Changsha, China}

\author{Chu-Wen Xiao}
\email{xiaochw@gxnu.edu.cn}
\affiliation{Department of Physics, Guangxi Normal University, Guilin 541004, China}
\affiliation{Guangxi Key Laboratory of Nuclear Physics and Technology, Guangxi Normal University, Guilin 541004, China}
\affiliation{School of Physics, Hunan Key Laboratory of Nanophotonics and Devices, Central South University, Changsha 410083, China}

\begin{abstract}

In present work, we study the reactions $D^{0} \rightarrow K^{+} K^{-} \eta$ and $D^0 \rightarrow \pi^{+} \pi^{-} \eta$, and find that the $a_0(980)$ state plays a dominant role. 
At the quark level, the external and internal $W$-emission mechanisms are taken into account, which can hadronize into the final states, and then the $a_0(980)$ state is generated from the final state interaction. 
Besides, the contributions of other intermediate resonances, such as $\rho(770)$ and $\phi(1020)$, are also considered. 
We make a combined fit of the invariant mass spectra measured by the Belle and BESIII Collaborations, where the results are in good agreement with the experiments, and the signal of the $a_0(980)$ shows great significance. 
Besides, the antisymmetry data for the production of the $a_0(980)^+$ and $a_0(980)^-$ is described well in the combined fit.

\end{abstract}

\maketitle


\section{Introduction}

Even though the states $a_0(980)$ and $f_0(980)$ have been found for a long time ago~\cite{Astier:1967zz,Ammar:1969vy,Defoix:1969qx}, the nature and structure of them remain topics of active research and debate in particle physics, see more discussions in the review of Particle Data Group (PDG)~\cite{ParticleDataGroup:2024cfk} for the meson resonances below 1 GeV. 
As already known that, they exhibit unique properties that challenge conventional understanding. 
The $a_0(980)$ is an isovector ($I=1$), while the $f_0(980)$ is an isoscalar ($I=0$), of which their masses are approximately 980 MeV and just below the $K\bar{K}$ threshold, leading to various interpretations concerning their properties. 
In the quark model, these resonances might be viewed as conventional scalar $q\bar{q}$ states~\cite{Godfrey:1985xj,Morgan:1993td,Tornqvist:1995ay}. 
However, the low masses of the $a_0(980)$ and $f_0(980)$ raise questions about their compatibility with this picture, particularly since the expected mass of the lightest scalar $q\bar{q}$ state, the $\sigma$ or $f_0(500)$, is significantly different. 
On the other hand, some thoeretical points of view suggested these particles could be tetraquarks, composed of four quarks ($qq\bar{q}\bar{q}$)~\cite{Jaffe:1976ig,Weinstein:1982gc,Achasov:2003cn}, which could explain some of their unusual properties and decay patterns.
An alternative and increasingly popular interpretation was that the $a_0(980)$ and $f_0(980)$ could be considered as molecular states, composed primarily of $K\bar{K}$ pairs~\cite{Zou:1994ea,Janssen:1994wn,Oller:1997ti,Locher:1997gr,Oller:1997ng}. 
This interpretation is supported by the proximity of their masses to the $K\bar{K}$ threshold and a strong coupling to the $K\bar{K}$ channel. 
Furthermore, some models proposed that the $a_0(980)$ and $f_0(980)$ could possess more complex structures involving contributions from other resonances, or even non-perturbative effects that lead to a more nuanced picture than a simple quark pair or a molecular model~\cite{Baru:2003qq,Albuquerque:2023bex}. 

Thus, many  experimental efforts, including the analysis of their decay modes, the scattering experiments, and resonant production mechanisms in various reactions, aim to clarify the nature of the $a_0(980)$ and $f_0(980)$. 
Moreover, the measurements of branching ratios in different decay modes, studying how these states couple to different final states and probing their production in high-energy collisions can help to distinguish between different theoretical frameworks. 
Their productions in the $\phi$ meson radiative decays were searched in the decays $\phi \to \gamma \pi^+\pi^-$~\cite{CMD-2:1999imm}, $\phi \to \gamma \pi^0\pi^0$~\cite{KLOE:2006vmv}, $\phi \to \gamma \pi^0\eta$~\cite{KLOE:2009ehb}, with the corresponding branching ratios measured. 
Analogous productions in the $J/\psi$ meson radiative decays were done in the decays $J/\psi \to \gamma \pi^0\pi^0$~\cite{BESIII:2015rug}, $J/\psi \to \gamma \pi^0\eta$~\cite{BESIII:2016gkg}, $J/\psi \to \gamma (\pi\pi,\, K\bar{K},\, \eta\eta,\, \omega\phi)$~\cite{Sarantsev:2021ein}, and so on, where the resonant signal was found. 
In the hadronic decay modes, the productions of the $a_0(980)$ and $f_0(980)$ can be easy in some certain decay processes. 
Three-body decays of charmed mesons, such as those involving $D$ mesons, represent a good chance for exploring low-lying scalar resonances, since the presence of three mesons in the final state provides a fertile ground for studying the properties of various resonances, including the $a_0(980)$ and $f_0(980)$ states.
The LHCb Collaboration employed a Dalitz plot method to analyze information about the resonance contributions, such as the $a_0(980)$, $f_0(980)$, $\phi(1020)$ and $f_0(1370)$ in the decay $D^+ \to K^-K^+K^+$~\cite{LHCb:2019tdw}, which was a doubly Cabibbo-suppressed process. 
Ref.~\cite{BESIII:2020pxp} reported the results of the first measurements of the absolute branching fractions for fourteen hadronic $D^{0(+)}$ decays with an $\eta$ in the exclusive final states, for example the ones $D^0 \to K^+K^-\eta,\, K^0_SK^0_S\eta$, and so on. 
Aim at hinting the $CP$ violation effect, the singly Cabibbo-suppressed decays $D^0 \to K^+K^-\eta,\, \pi^+\pi^-\eta,\, \phi \eta$ were investigated in Ref.~\cite{Belle:2021dfa}, where the branching fractions and $CP$ asymmetries were measured, and the invariant mass distributions were also given with the resonance signals, such as the $a_0(980)$, appearing. 
The $CP$ asymmetries were also tested in the decays $D^0 \to \pi^+\pi^-\eta$ and $D^+ \to \pi^+\pi^0\eta$ with the absolute branching fractions measured~\cite{BESIII:2019xhl}, where no $CP$ violation was found and which were measured first time by the CLEO Collaboration~\cite{CLEO:2008icw}. 
Recently, the BESIII Collaboration reported the updated results of the first amplitude analyses for the decays $D^{0(+)} \to \pi^+\pi^{-(0)}\eta$ in Ref.~\cite{BESIII:2024tpv}, where they found that the production of $a_0(980)^+$ was much larger than the one of  $a_0(980)^{-(0)}$ in the intermediate processes of these two decays, indicating final state isospin symmetry breaking, and the contribution of the $\rho^{0(+)}$ resonances was important. 
Furthermore, the $a_0(980)$ and $f_0(980)$ can also be reproduced in the certain final state interactions of three-body hadronic decays of the $D_s$, $B_{(s)}$ mesons, of course some other $D$ meson decay processes too, see more results in PDG~\cite{ParticleDataGroup:2024cfk}.

On the theoretical side, there are also many works with different models or approaches devoting great effort to shed light on the dynamics of strong interactions and the resonance nature. 
Using a covariant quark model, the properties of the $f_0(980)$ resonance were investigated in Ref.~\cite{El-Bennich:2008rkp} in the decays $D_{(s)} \to f_0(980) \pi,\, f_0(980) K$. 
Within the perturbative QCD approach, Ref.~\cite{Yang:2021zcx} systematically analyzed the charmless decays $B_{(s)} \to V \pi\pi$ with $V$ representing vector mesons, where the branching fractions were evaluated by considering the $f_0(980)$ resonance contribution. 
In Ref.~\cite{Aoude:2018zty}, the decay $D^+ \to K^-K^+K^+$ was investigated using the isobar model alongside the coupled-channel $K$-matrix approach, allowing for a detailed examination of the $a_0(980)$ and $f_0(980)$ resonance contributions. 
This double Cabibbo-suppressed decay, $D^+ \to K^-K^+K^+$, was also examined by exploring the $N/D$ method in Ref.~\cite{Roca:2020lyi}, where the data of the invariant mass distributions was fitted well. 
Applying a coupled-channel framework that accounted for the final state interactions, Ref.~\cite{Nakamura:2015qga} analyzed the decay $D^+ \to K^-\pi^+\pi^+$, which was also studied in Refs.~\cite{Niecknig:2015ija,Niecknig:2017ylb} with the dispersion theory based on the Khuri-Treiman formalism. 
Furthermore, based on the final state interaction from the chiral unitary approach (ChUA)~\cite{Oller:1997ti,Oset:1997it,Kaiser:1998fi,Oller:2000fj}, the results of Ref.~\cite{Xie:2014tma} indicated that the resonance contributions from the $f_0(500)$, $f_0(980)$ and $a_0(980)$ in  the $\pi^+\pi^-$ and $\pi^0\eta$ invariant mass distributions of the decays $D^0 \to \bar{K}^0\pi^+\pi^-$ and $D^0 \to \bar{K}^0\pi^0\eta$, respectively, were important, where a obtained ratio of the branching fractions was consistent with experimental measurement within the uncertainties. 
With similar formalism, the singly Cabibbo-suppressed processes $D^0 \to \pi^0\pi^0\pi^0,\, \pi^0\pi^0\eta,\, \pi^0\eta\eta$ were studied in Ref.~\cite{Wang:2021kka}, where the resonances $f_0(500)$, $f_0(980)$ and $a_0(980)$ were dynamically reproduced in the final state interactions based on the ChUA, and the data of the $\pi^0\eta$ invariant mass spectrum of the decay $D^0 \to \pi^0\eta\eta$, measured by the BESIII Collaboration~\cite{BESIII:2018hui}, was described well. 

In the present work, aiming at understanding the production properties of the $a_0(980)$ and $f_0(980)$, we analyze the decays $D^0 \to K^+K^-\eta$ and $D^0 \to \pi^+\pi^-\eta$ based on the experimental measurements of Refs.~\cite{Belle:2021dfa,BESIII:2024tpv} applying the final state interaction framework within the ChUA. Our manuscript is organized as following. In the next section, our formalism of the final state interaction is presented. And then, the results for the decays $D^0 \to K^+K^-\eta$ and $D^0 \to \pi^+\pi^-\eta$ are discussed in detail. Finally, it is a short summary.

\section{Formalism}\label{Sec2}

In the present work,  we investigate the weak decay process of the $D^{0} \rightarrow K^{+} K^{-} \eta$ and $\pi^{+} \pi^{-} \eta$ reactions based on the ChUA for the final state interaction. 
First, the resonance $a_0(980)$ is dynamically generated by the $S$-wave pseudoscalar-pseudoscalar meson interaction. 
Then, the states $\phi$ and $\rho$ are contributed from the $P$-wave, which are calculated with the Breit-Wigner amplitudes as done in Refs.~\cite{Toledo:2020zxj,Wang:2021ews,Liang:2023ekj}. 
Finally, the double differential widths of the $D^{0} \rightarrow K^{+} K^{-} \eta$ and $\pi^{+} \pi^{-} \eta$ processes are evaluated and fitted to the experimental data.

\begin{figure}[!htbp]
\centering
\includegraphics[width=1\linewidth]{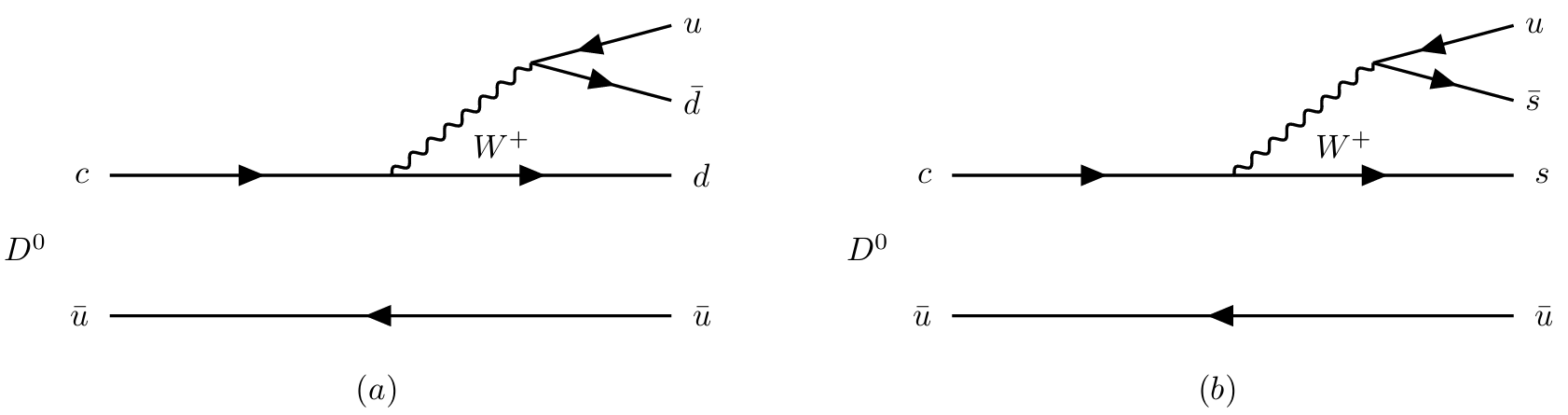}
\caption{External $W$-emission mechanism for the processes (a) $c \to W^+ d$; (b) $c \to W^+ s$.} \label{external}
\end{figure}

\begin{figure}[!htbp]
\centering
\includegraphics[width=1\linewidth]{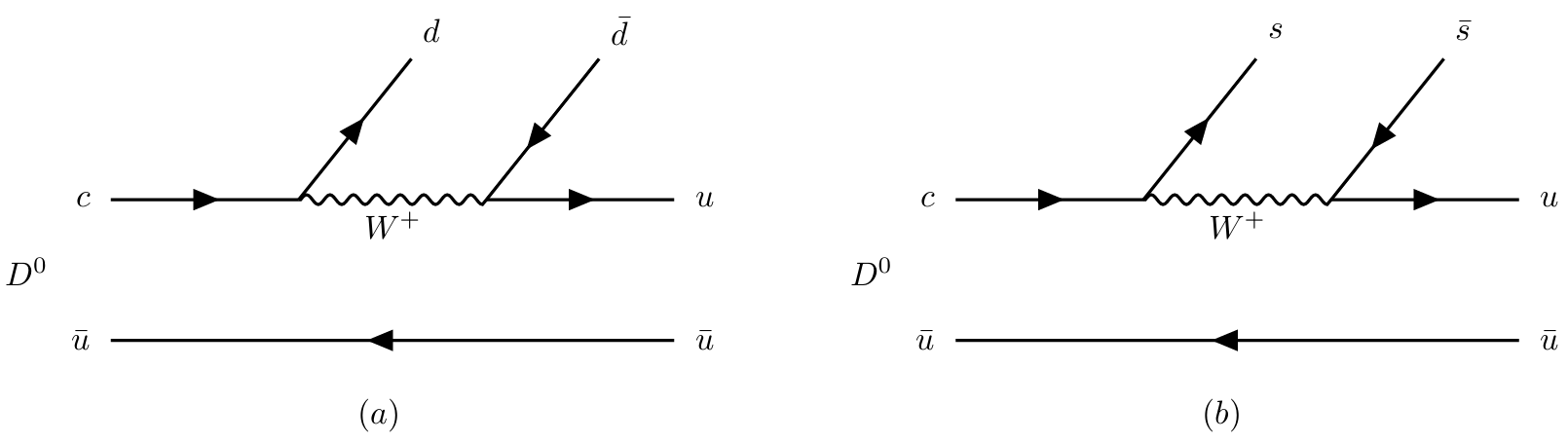}
\caption{Internal $W$-emission mechanism for the processes (a) $c \to W^+ d$; (b) $c \to W^+ s$.} \label{internal}
\end{figure}

The contributions of weak decay topography have several cases~\cite{Chau:1982da,Chau:1987tk,Morrison:1989xq,Molina:2019udw}, and only the dominant mechanisms for the external and internal $W$-emission are considered, which are depicted in Figs.~\ref{external} and \ref{internal}. Note that these diagrams can all contribute to the two decays that we concern. 
For the $W$-external emission at the quark level, as shown in Fig.~\ref{external}, the quark $c$ of the initial meson $D^0$ can decay into $W^{+}$ boson and $d(s)$ quark, while the $\bar{u}$ quark of the $D^0$ meson remains as a spectator. 
Then, there are two cases for the hadronization procedures. 
The first one is that, the quark pair $u \bar{d}(u \bar{s})$ created by the $W$-boson can form the $\pi^+(K^+)$ meson directly, and the other pairs $d \bar{u}(s \bar{u})$ generates two final states through the hadronization with quark pairs $q\bar{q}=u\bar{u}+d\bar{d}+s\bar{s}$ produced from the vacuum. 
The second one is that, the quark pairs $d\bar{u}(s\bar{u})$ becomes the $\pi^-(K^-)$, and the other quark pair $u\bar{d}(u\bar{s})$ undergoes the hadronization. 
Then, the corresponding decay procedures for these hadronizations can be written as 
\begin{equation}\label{eq1}
\begin{aligned}
H^{(1a)}  &= V_{P}V_{cd}V_{ud} \lbrace \left( u\bar{d}\rightarrow \pi
^{+}\right) \left[ d\bar{u}\rightarrow d\bar{u}\left( u\bar{u}+d\bar{d}+s%
\bar{s}\right) \right] \\
&\quad + \left( d\bar{u}\rightarrow \pi ^{-}\right) \left[ u%
\bar{d}\rightarrow u\bar{d}\left( u\bar{u}+d\bar{d}+s\bar{s}\right) \right] %
\rbrace \\  
&= V_{P}V_{cd}V_{ud} \lbrace \left( u\bar{d}\rightarrow \pi ^{+}\right) %
\left[ M_{21}\rightarrow \left( M\cdot M\right) _{21}\right]  \\
&\quad +\left( d\bar{u}%
\rightarrow \pi ^{-}\right) \left[ \left( M\cdot M\right) _{12}\right] %
\rbrace,
\end{aligned}
\end{equation}
\begin{equation}\label{eq2}
\begin{aligned}
H^{(1b)} & = V_{P}^{\prime}V_{cs}V_{us}\left( u\bar{s}\rightarrow K^{+}\right) 
\lbrace s\bar{u}\rightarrow s\bar{u}\left( u\bar{u}+d\bar{d}+s\bar{s}\right) \\
&\quad +\left( s\bar{u}\rightarrow K^{-}\right) \left[ u\bar{s}\rightarrow u\bar{s}%
\left( u\bar{u}+d\bar{d}+s\bar{s}\right) \right] \rbrace \\
& = V_{P}^{\prime}V_{cs}V_{us} \lbrace \left( u\bar{s}\rightarrow K^{+}\right) 
\left[ M_{31}\rightarrow \left( M\cdot M\right) _{31}\right] \\
&\quad +\left( s\bar{u}%
\rightarrow K^{-}\right) \left[ M_{13}\rightarrow \left( M\cdot M\right)
_{13}\right] \rbrace.
\end{aligned}
\end{equation}

The $W$-internal emission mechanism are shown in Fig.~\ref{internal}. 
Similarly, the quark $c$ of the initial meson $D^0$ can decay into $W^{+}$ boson and $d(s)$ quark, while the $\bar{u}$ quark of the $D^0$ meson remains as a spectator. 
But, the two cases of hadronization processes occur between two quark pairs of $u\bar{u}$ and $d\bar{d} (s\bar{s})$. 
Analogously, these hadronization processes can be written as 
\begin{equation}\label{eq3}
\begin{aligned}
H^{(2a)} &= \beta V_{P} V_{cd}V_{ud}\left[\left( d\bar{d}\rightarrow -\frac{1}{\sqrt{2}}\pi ^{0}\right) \left[ u\bar{u} \rightarrow u\bar{u}\left( u\bar{u}+d\bar{d}+s\bar{s}\right) \right] \right. \\
&\quad \left.+\left( d\bar{d}\rightarrow \frac{1}{\sqrt{6}}\eta \right) \left[ u\bar{u} \rightarrow u\bar{u}\left( u\bar{u}+d\bar{d}+s\bar{s}\right) \right] \right. \\
&\quad \left.+\left( u\bar{u}\rightarrow \frac{1}{\sqrt{2}}\pi ^{0}\right) \left[ d\bar{d} \rightarrow d\bar{d}\left( u\bar{u}+d\bar{d}+s\bar{s}\right) \right] \right. \\
&\quad \left.+\left( u\bar{u}\rightarrow \frac{1}{\sqrt{6}}\eta \right) \left[
d\bar{d} \rightarrow d\bar{d}\left( u\bar{u}+d\bar{d}+s\bar{s}\right) \right] \right]  \\
&=  \beta V_{P} V_{cd}V_{ud} \left[\left( d\bar{d}\rightarrow -\frac{1}{\sqrt{2}}\pi ^{0}\right) \left[M_{11}\rightarrow \left( M\cdot M\right) _{11}\right] \right. \\
&\quad \left.+\left( d\bar{d}\rightarrow \frac{1}{\sqrt{6}}\eta \right) \left[
M_{11}\rightarrow \left( M\cdot M\right) _{11}\right] \right. \\
&\quad \left.+\left( u\bar{u}\rightarrow \frac{1}{\sqrt{2}}\pi ^{0}\right) \left[
M_{22}\rightarrow \left( M\cdot M\right) _{22}\right] \right. \\
&\quad \left.+\left( u\bar{u}\rightarrow \frac{1}{\sqrt{6}}\eta \right) \left[
M_{22}\rightarrow \left( M\cdot M\right) _{22}\right] \right],  \\
\end{aligned}
\end{equation}
\begin{equation}\label{eq4}
\begin{aligned}
H^{(2b)} &= \beta V_{P}^{\prime} V_{cs}V_{us} \left[ \left( s\bar{s}\rightarrow -\frac{2}{\sqrt{6}}\eta \right) \left[ u\bar{u} \rightarrow u\bar{u}\left( u\bar{u}+d\bar{d}+s\bar{s}\right) \right] \right.  \\
&\quad \left.+\left( u\bar{u}\rightarrow \frac{1}{\sqrt{2}}\pi ^{0}\right) \left[ s\bar{s} \rightarrow s\bar{s}\left( u\bar{u}+d\bar{d}+s\bar{s}\right) \right] \right. \\
&\quad \left. +\left( u\bar{u}\rightarrow \frac{1}{\sqrt{6}}\eta \right) \left[ s\bar{s} \rightarrow s\bar{s}\left( u\bar{u}+d\bar{d}+s\bar{s}\right) \right] \right]  \\
&= \beta V_{P}^{\prime} V_{cs}V_{us}\left[ \left( s\bar{s}\rightarrow -\frac{2}{\sqrt{6}}\eta \right) \left[ M_{11}\rightarrow \left( M\cdot M\right) _{11}\right] \right. \\
&\quad \left. +\left( u\bar{u}\rightarrow \frac{1}{\sqrt{2}}\pi ^{0}\right) \left[ M_{33}\rightarrow \left( M\cdot M\right) _{33} \right] \right.  \\
&\quad \left. +\left( u\bar{u}\rightarrow \frac{1}{\sqrt{6}}\eta \right) \left[
M_{33}\rightarrow \left( M\cdot M\right) _{33}\right] \right], \\
\end{aligned}
\end{equation}
where $V_{P}$ and $V_{P}^{\prime}$ are the vertex factors of the weak interaction strength, which are assumed as constants~\cite{Wang:2021ews,Ahmed:2020qkv,Wang:2020pem,Peng:2024ive}. 
$\beta$ is the relative weight coefficient between the internal and external $W$-emission mechanisms~\cite{Zhang:2022xpf,Dai:2018nmw}. 
The factors $-1/\sqrt{2}$, $1/\sqrt{2}$, $1/\sqrt{6}$ and $-2/\sqrt{6}$ come from the flavour components of $\pi^0$ and $\eta$, which are given as $\left\vert \pi ^{0}\right\rangle =\frac{1}{\sqrt{2}}\left\vert u\bar{u}-d \bar{d}\right\rangle $, $\left\vert \eta \right\rangle = \frac{1}{\sqrt{6}}\left\vert u\bar{u}+d\bar{d}-2s\bar{s}\right\rangle$. 
The $V_{q_1 q_2}$ is the element of the Cabibbo-Kabayashi-Maskawa(CKM) matrix for the transition of the quark $q_1 \rightarrow q_2$, and the matrix $M$ is the $SU(3)$ matrix of the pseudoscalar mesons, given by
\begin{eqnarray}
M =  \left( 
\begin{array}{ccc}
u\bar{u} & u\bar{d} & u\bar{s} \\ 
d\bar{u} & d\bar{d} & d\bar{s} \\ 
s\bar{u} & s\bar{d} & s\bar{s}%
\end{array}%
\right) 
 =  \left( 
\begin{array}{ccc}
\frac{1}{\sqrt{2}}\pi ^{0}+\frac{1}{\sqrt{6}}\eta & \pi ^{+} & K^{+} \\ 
\pi ^{-} & -\frac{1}{\sqrt{2}}\pi ^{0}+\frac{1}{\sqrt{6}}\eta & K^{0} \\ 
K^{-} & \bar{K}^{0} & -\frac{2}{\sqrt{6}}\eta%
\end{array}%
\right), 
\end{eqnarray}
where we take $\eta \equiv \eta_8$ as done in Ref.~\cite{Toledo:2020zxj}. 
Then, the hadronization processes at the quark level in Eqs.~(\ref{eq1})-(\ref{eq4}) can be rewritten at the hadron level as
\begin{eqnarray}
\begin{aligned}
\left( M\cdot M\right) _{11} &= \frac{1}{2}\pi ^{0}\pi ^{0}+\frac{1}{6}\eta \eta +\frac{1}{\sqrt{3}}\pi^{0}\eta +\pi ^{+}\pi ^{-}+K^{+}K^{-}, \\ 
\left( M\cdot M\right) _{12} &= \frac{2}{\sqrt{6}}\pi ^{+}\eta +K^{+}\bar{K}^{0}, \\
\left( M\cdot M\right) _{13} &= \frac{1}{\sqrt{2}}\pi ^{0}K^{+}-\frac{1}{\sqrt{6}}K^{+}\eta +\pi ^{+}K^{0}, \\
\left( M\cdot M\right) _{21} &= \frac{2}{\sqrt{6}}\pi ^{-}\eta +K^{0}K^{-},  \\
\left( M\cdot M\right) _{22} &= \pi ^{+}\pi ^{-}+\frac{1}{2}\pi ^{0}\pi ^{0}+\frac{1}{6}\eta \eta -\frac{1}{\sqrt{3}}\pi ^{0}\eta +K^{0}\bar{K}^{0},  \\
\left( M\cdot M\right) _{31} &= \frac{1}{\sqrt{2}}K^{-}\pi ^{0}-\frac{1}{\sqrt{6}}\eta K^{-}+\bar{K}^{0}\pi ^{-},   \\
\left( M\cdot M\right) _{33} &= K^{+}K^{-}+K^{0}\bar{K}^{0}+\frac{2}{3}\eta \eta .
\end{aligned}
\end{eqnarray}%

Therefore, one can get all the possible final states with $\pi^+(\pi^-)$, $K^+(K^-)$, $\pi^0$ and $\eta$ directly produced after the hadronization,
\begin{eqnarray}
H^{\left( 1a\right) } &= V_{P}V_{cd}V_{ud}\left( \frac{4}{\sqrt{6}}\pi ^{+}\pi ^{-}\eta +\pi^{+}K^{0}K^{-}+\pi ^{-}K^{+}\bar{K}^{0}\right),
\end{eqnarray}
\begin{eqnarray}
H^{\left( 1b\right) } &=V_{P}^{^{\prime }}V_{cs}V_{us}\left( \frac{2}{\sqrt{2}}K^{+}K^{-}\pi ^{0}- \frac{2}{\sqrt{6}}\eta K^{+}K^{-}+K^{+}\bar{K}^{0}\pi ^{-}+\pi ^{+}K^{0}K^{-} \right),
\end{eqnarray}
\begin{equation}
\begin{aligned}
H^{(2 a)}&=  \beta V_{P}V_{cd}V_{ud}\left( \frac{1}{3\sqrt{6}}\eta \eta \eta -\frac{1}{\sqrt{2}}\pi ^{0}K^{+}K^{-}+\frac{2}{\sqrt{6}}\pi ^{+}\pi ^{-}\eta +\frac{1}{\sqrt{6}}K^{+}K^{-}\eta \right. \\
&\quad \left.-\frac{1}{\sqrt{6}}\pi ^{0}\pi ^{0}\eta +\frac{1}{\sqrt{2}}\pi ^{0}K^{0}\bar{%
K}^{0}+\frac{1}{\sqrt{6}}K^{0}\bar{K}^{0}\eta \right),
\end{aligned}
\end{equation}
\begin{equation}
\begin{aligned}
H^{(2 b)}&=  \beta V_{P}^{^{\prime }}V_{cs}V_{us}\left(  -\frac{1}{\sqrt{6}}\pi ^{0}\pi ^{0}\eta +\frac{1}{3\sqrt{6}}\eta \eta \eta - \frac{2}{\sqrt{6}}\pi ^{+}\pi ^{-}\eta -\frac{1}{\sqrt{6}}K^{+}K^{-}\eta \right. \\
&\quad \left. +\frac{1}{\sqrt{2}}K^{+}K^{-}\pi ^{0}+\frac{1}{\sqrt{2}}\pi ^{0}K^{0}\bar{K}%
^{0}+\frac{1}{\sqrt{6}}K^{0}\bar{K}^{0}\eta \right).
\end{aligned}
\end{equation}
Due to the elements of the CKM matrix has the following relationship, $ V_{cd}V_{ud}=-V_{us}V_{cs}$~\cite{ParticleDataGroup:2024cfk}. 
Thus, we can obtain the total contributions in the $S$-wave of the weak decay processes $D^{0} \rightarrow \pi^{+} \pi^{-} \eta$ and $K^{+} K^{-} \eta$, having
\begin{equation}\label{eq11}
\begin{aligned}
H &=  H^{\left( 1a\right) }+H^{\left( 1b\right) }+H^{\left( 2a\right)
}+H^{\left( 2b\right) } \\
&= \frac{2}{\sqrt{6}}\left( 2V_{P}+\beta V_{P}+\beta V_{P}^{^{\prime }}\right)
V_{cd}V_{ud}\pi ^{+}\pi ^{-}\eta +\left( V_{P}-V_{P}^{^{\prime }}\right)
V_{cd}V_{ud}\pi ^{+}K^{0}K^{-} \\
&\quad + \left( V_{P}-V_{P}^{^{\prime }}\right)
V_{cd}V_{ud}\pi ^{-}K^{+}\bar{K}^{0} - \frac{1}{\sqrt{2}}\left( 2V_{P}^{^{\prime }}+\beta V_{P}+\beta V_{P}^{^{\prime }}\right) V_{cd}V_{ud}K^{+}K^{-}\pi ^{0} \\
&\quad +\frac{1}{\sqrt{6}}\left( 2V_{P}^{^{\prime }} + \beta V_{P} + \beta V_{P}^{^{\prime }}\right)V_{cd}V_{ud}K^{+}K^{-}\eta + \frac{1}{3\sqrt{6}}\beta \left( V_{P}-V_{P}^{^{\prime }}\right)V_{cd}V_{ud}\eta \eta \eta \\
&\quad +\frac{1}{\sqrt{6}}\beta \left( V_{P}^{^{\prime}}-V_{P}\right) V_{cd}V_{ud}\pi ^{0}\pi ^{0}\eta + \frac{1}{\sqrt{2}}\beta \left( V_{P}-V_{P}^{^{\prime }}\right)
V_{cd}V_{ud}\pi ^{0}K^{0}\bar{K}^{0} \\
&\quad +\frac{1}{\sqrt{6}}\beta \left(
V_{P}-V_{P}^{^{\prime }}\right) V_{cd}V_{ud}K^{0}\bar{K}^{0}\eta  \\
&= \frac{2}{\sqrt{6}}(2C_{1}+\beta C_{1}+\beta C_{2})\pi ^{+}\pi ^{-}\eta + (C_{1}-C_{2})\pi^{+}K^{0}K^{-} + (C_{1}-C_{2})\pi ^{-}K^{+}\bar{K}^{0} \\
&\quad -\frac{1}{\sqrt{2}}(2C_{2}+\beta C_{1}+\beta C_{2})K^{+}K^{-}\pi ^{0}+\frac{1}{\sqrt{6}}(2C_{2}+\beta C_{1}+\beta C_{2})K^{+}K^{-}\eta \\ 
&\quad + \frac{1}{3\sqrt{6}}\beta (C_{1}-C_{2})\eta \eta \eta +\frac{1}{\sqrt{6}}\beta
(C_{2}-C_{1})\pi ^{0}\pi ^{0}\eta +\frac{1}{\sqrt{2}}\beta (C_{1}-C_{2})\pi ^{0}K^{0}\bar{K}^{0} \\
&\quad +\frac{1}{\sqrt{6}}\beta (C_{1}-C_{2})K^{0}\bar{K}^{0}\eta, 
\end{aligned}
\end{equation}
where the factors $C_1$ and $C_2$ are defined as $C_{1} =  V_{P} V_{cd}V_{ud} $ and $C_{2} = V_{P}^{\prime } V_{cd}V_{ud}$, which are also absorbed the normalization factor of the data events and determined by fitting the experimental data later. 

\begin{figure}[htbp]
\begin{subfigure}{0.475\textwidth}
\centering
\includegraphics[width=1\linewidth]{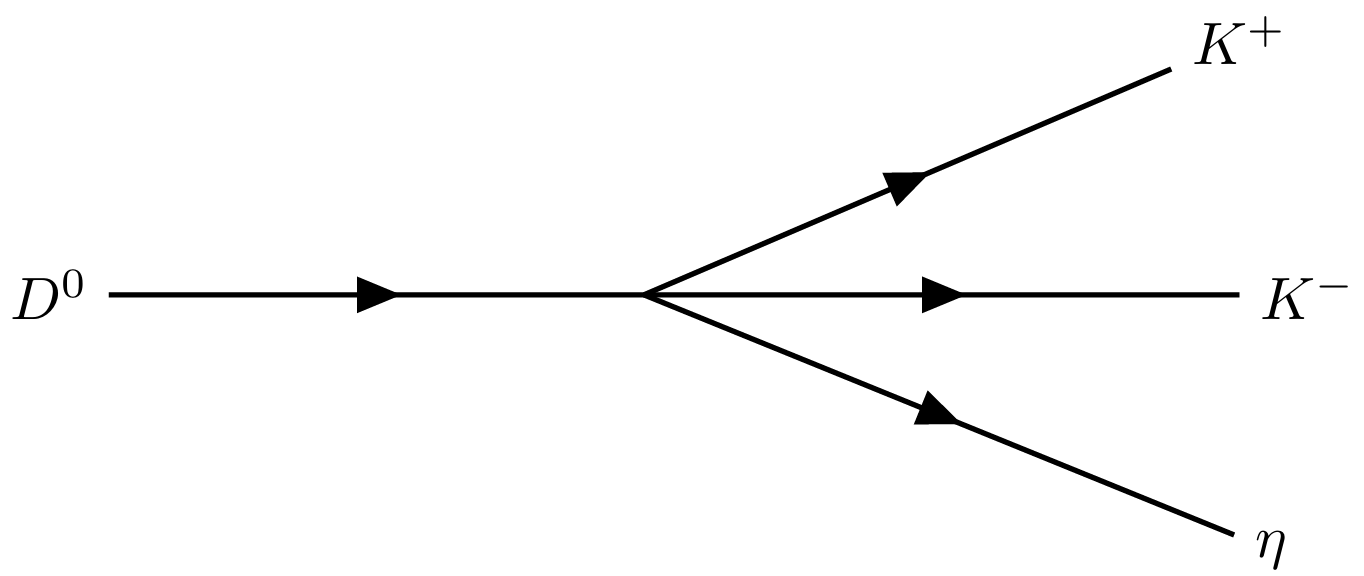} 
\caption{Tree-level production.}
\end{subfigure}
\begin{subfigure}{0.475\textwidth}  
\centering 
\includegraphics[width=1\linewidth]{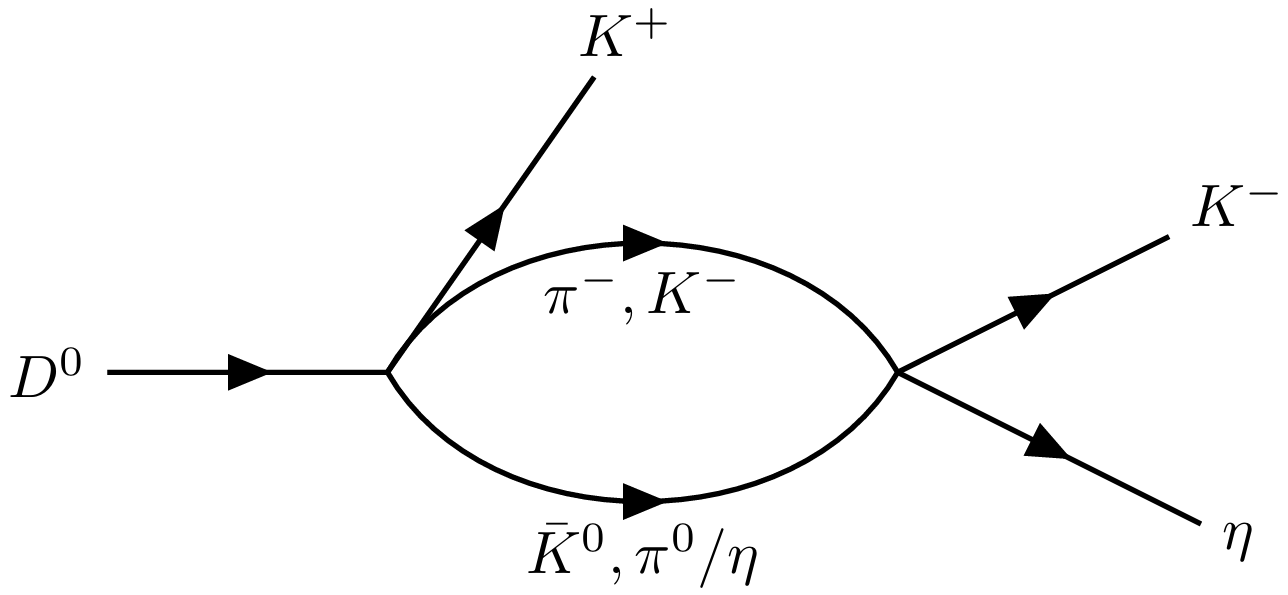} 
\caption{Rescattering of $\pi^- \bar{K}^0$, $K^- \pi^0$ and $K^- \eta$.}
\end{subfigure}	
\begin{subfigure}{0.475\textwidth}
\centering
\includegraphics[width=1\linewidth]{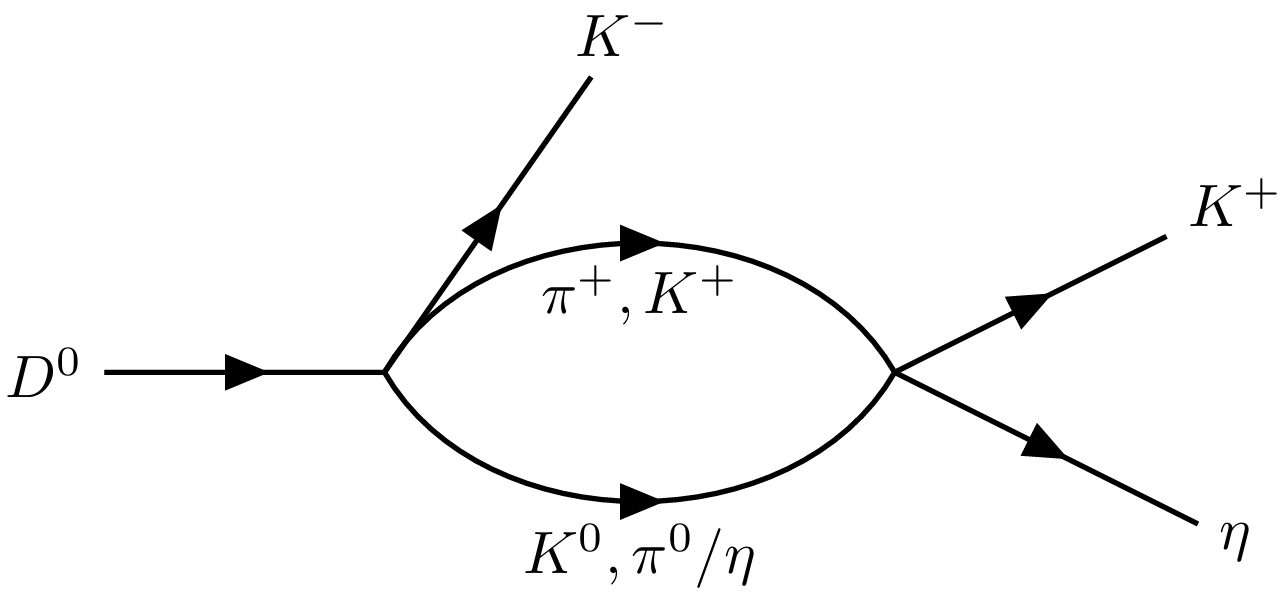} 
\caption{Rescattering of $\pi^+ K^0$, $K^+ \pi^0$ and $K^+ \eta$.}
\end{subfigure}
\begin{subfigure}{0.475\textwidth}  
\centering 
\includegraphics[width=1\linewidth]{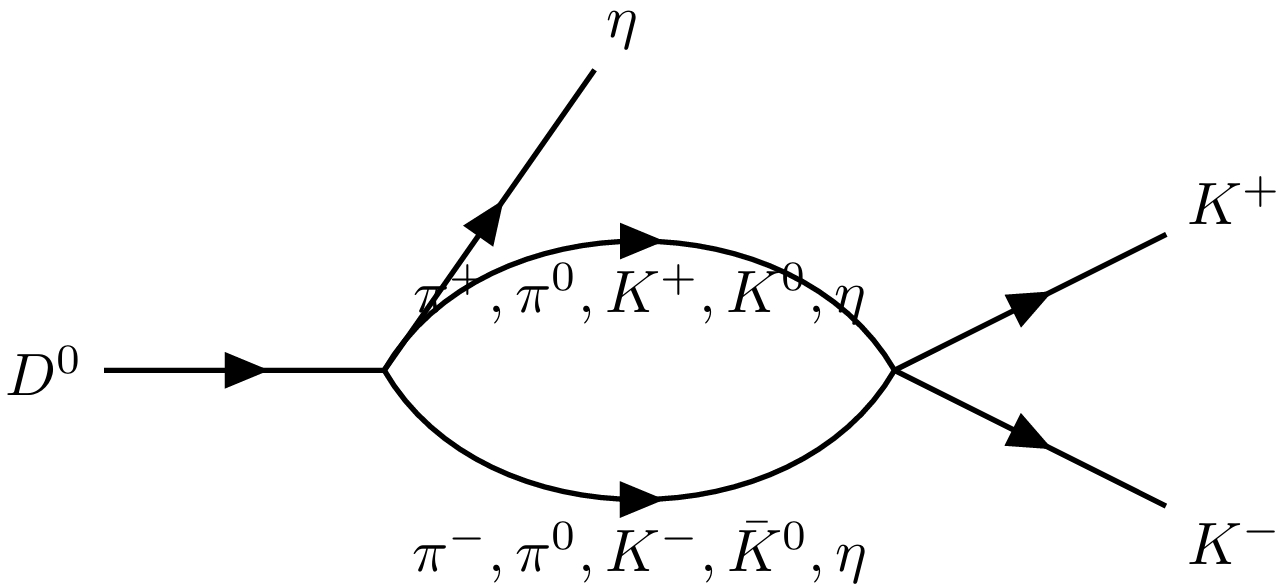} 
\caption{Rescattering of $\pi^+ \pi^-$, $\pi^0 \pi^0$, $K^+ K^-$, $K^0 \bar{K}^0$, and $\eta \eta$.}
\end{subfigure}	
\captionsetup{justification=raggedright}
\caption{Diagrammatic representations of rescattering  for the decay $D^{0}\rightarrow K^{+}K^{-}\eta$.}
\label{ScatterKKeta} 
\end{figure}

\begin{figure}[htbp]
\begin{subfigure}{0.475\textwidth}
\centering
\includegraphics[width=1\linewidth]{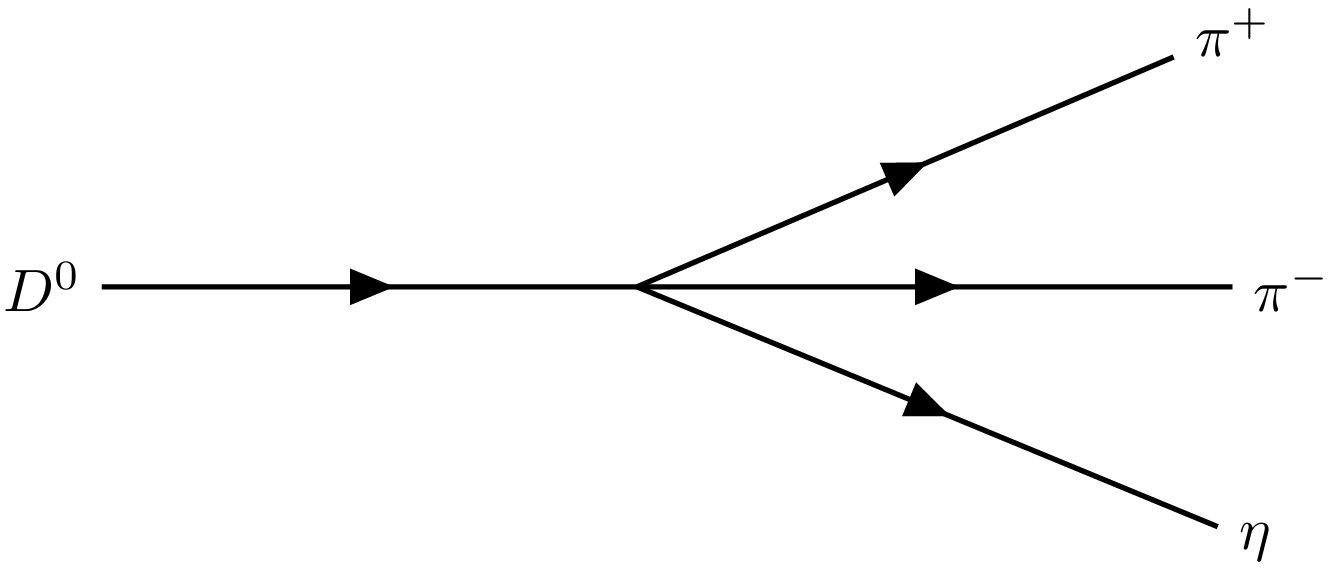} 
\caption{Tree-level production.}
\end{subfigure}
\begin{subfigure}{0.475\textwidth}  
\centering 
\includegraphics[width=1\linewidth]{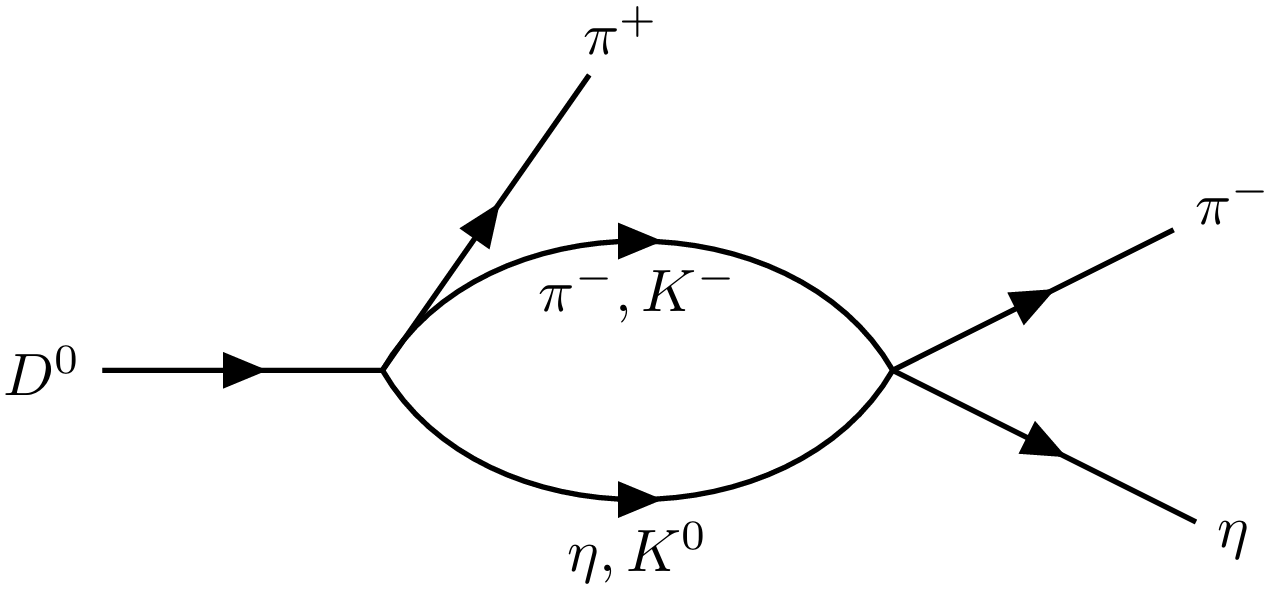} 
\caption{Rescattering of $\pi^- \eta$ and $K^- K^0$.}
\end{subfigure}	
\begin{subfigure}{0.475\textwidth}
\centering
\includegraphics[width=1\linewidth]{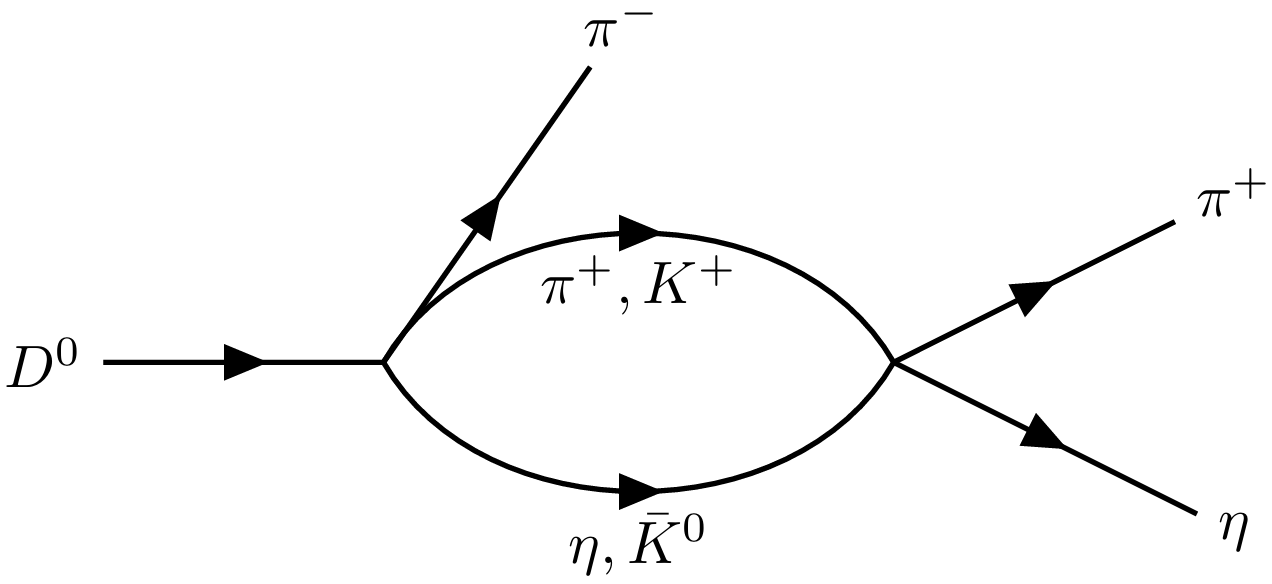} 
\caption{Rescattering of $\pi^+ \eta$ and $K^+ \bar{K}^0$.}
\end{subfigure}
\begin{subfigure}{0.475\textwidth}  
\centering 
\includegraphics[width=1\linewidth]{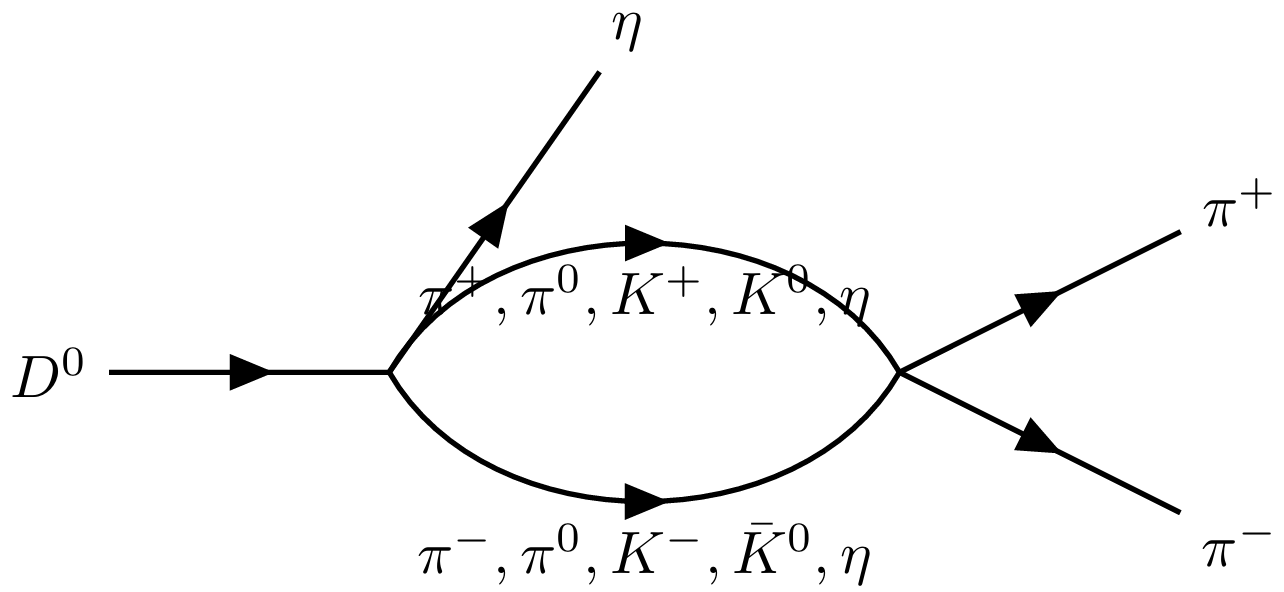} 
\caption{Rescattering of $\pi^+ \pi^-$, $\pi^0 \pi^0$, $K^+ K^-$, $K^0 \bar{K}^0$, and $\eta \eta$.}
\end{subfigure}	
\captionsetup{justification=raggedright}
\caption{Diagrammatic representations of rescattering for the decay $D^{0}\rightarrow \pi^+ \pi^-\eta$.}
\label{Scatterpipieta}
\end{figure}

Note that, the terms $K^+ K^- \eta$ and $\pi^+ \pi^- \eta$ not only can obtain from the final states on the tree-level hadronization processes, but also produce through the rescattering procedures, as depicted in Figs.~\ref{ScatterKKeta} and \ref{Scatterpipieta}.
Thus, after considering the rescattering processes, the full amplitude for the decays $D^0 \rightarrow K^+ K^- \eta / \pi^+ \pi^- \eta$ in $S$ wave can be obtained as follows,
\begin{equation}\label{eq12}
\begin{aligned}
t_{D^{0}\rightarrow K^{+}K^{-}\eta } (s_{12}, s_{23}) &=   \frac{2}{\sqrt{6}} (2C_{1}+\beta C_{1}+\beta C_{2}) G_{\pi^{+}\pi ^{-}}\left( s_{12}\right) T_{\pi ^{+}\pi ^{-}\rightarrow
K^{+}K^{-}}\left( s_{12}\right) +\frac{1}{\sqrt{6}}\left(2C_{2}+\beta C_{1}+\beta C_{2}\right) \\
&\quad + \frac{1}{\sqrt{6}} \left(2C_{2}+\beta C_{1}+\beta C_{2} \right) G_{K^{+}K^{-}}\left( s_{12}\right) T_{K^{+}K^{-} \rightarrow K^{+}K^{-}}\left( s_{12}\right)  \\
&\quad + \frac{1}{3\sqrt{6}}\beta (C_{1}-C_{2}) G_{\eta \eta }\left( s_{12}\right) T_{\eta
\eta \rightarrow K^{+}K^{-}}\left( s_{12}\right) \\
&\quad  -\frac{1}{\sqrt{6}}\beta (C_{1}-C_{2}) G_{\pi ^{0}\pi ^{0}}\left( s_{12}\right) T_{\pi ^{0}\pi ^{0}\rightarrow
K^{+}K^{-}}\left( s_{12}\right)  \\
&\quad + \frac{1}{\sqrt{6}} \beta (C_{1}-C_{2}) G_{K^{0}\bar{K}^{0}}\left( s_{12}\right)
T_{K^{0}\bar{K}^{0}\rightarrow K^{+}K^{-}}\left( s_{12}\right) \\
&\quad +(C_{1}-C_{2}) G_{\pi^{-}\bar{K}^{0}}\left( s_{23}\right) T_{\pi ^{-}\bar{K}^{0}\rightarrow K^{-}\eta }\left( s_{23}\right)  \\
&\quad - \frac{1}{\sqrt{2}} \left( 2C_{2}+\beta C_{1}+\beta C_{2}\right) G_{K^{-}\pi ^{0}}\left(
s_{23}\right) T_{K^{-}\pi ^{0}\rightarrow K^{-}\eta }\left( s_{23}\right) \\
&\quad + \frac{1}{\sqrt{6}} \left( 2C_{2}+\beta C_{1}+\beta C_{2}\right) G_{K^{-}\eta }\left(
s_{23}\right) T_{K^{-}\eta \rightarrow K^{-}\eta }\left( s_{23}\right)  \\
&\quad + (C_{1}-C_{2}) G_{\pi ^{+}K^{0}}\left( s_{13}\right) T_{\pi ^{+}K^{0}\rightarrow
K^{+}\eta }\left( s_{13}\right) \\
&\quad -\frac{1}{\sqrt{2}} \left( 2C_{2}+\beta C_{1}+\beta C_{2} \right) G_{K^{+}\pi ^{0}}\left( s_{13}\right) T_{K^{+}\pi^{0}\rightarrow K^{+}\eta }\left( s_{13}\right)  \\
&\quad + \frac{1}{\sqrt{6}} \left( 2C_{2}+\beta C_{1}+\beta C_{2} \right) G_{K^{+}\eta }\left(
s_{13}\right) T_{K^{+}\eta \rightarrow K^{+}\eta }\left( s_{13}\right),   \\
\end{aligned}
\end{equation}\begin{equation}\label{eq13}
\begin{aligned}
t_{D^{0}\rightarrow \pi ^{+}\pi ^{-}\eta } (s_{12}, s_{23}) & = \frac{2}{\sqrt{6}} (2C_{1}+\beta C_{1}+\beta C_{2}) +\frac{2}{\sqrt{6}} (2C_{1}+\beta C_{1}+\beta C_{2}) G_{\pi ^{+}\pi ^{-}}\left( s_{12}\right) T_{\pi ^{+}\pi
^{-}\rightarrow \pi ^{+}\pi ^{-}}\left( s_{12}\right) \\
&\quad + \frac{1}{\sqrt{6}} \left( 2C_{2}+\beta C_{1}+\beta C_{2} \right) G_{K^{+}K^{-}}\left( s_{12}\right) T_{K^{+}K^{-}\rightarrow \pi ^{+}\pi ^{-}}\left( s_{12}\right)  \\
&\quad + \frac{1}{3\sqrt{6}} \beta (C_{1}-C_{2})G_{\eta \eta }\left( s_{12}\right) T_{\eta
\eta \rightarrow \pi ^{+}\pi ^{-}}\left( s_{12}\right) \\
&\quad - \frac{1}{\sqrt{6}} \beta (C_{1}-C_{2})G_{\pi ^{0}\pi ^{0}}\left( s_{12}\right) T_{\pi ^{0}\pi^{0}\rightarrow \pi ^{+}\pi ^{-}}\left( s_{12}\right) \\
&\quad + \frac{1}{\sqrt{6}}\beta (C_{1}-C_{2})G_{K^{0}\bar{K}^{0}}\left( s_{12}\right)
T_{K^{0}\bar{K}^{0}\rightarrow \pi ^{+}\pi ^{-}}\left( s_{12}\right)  \\
&\quad + \frac{2}{\sqrt{6}} (2C_{1}+\beta C_{1}+\beta C_{2}) G_{\pi ^{-}\eta }\left( s_{23}\right) T_{\pi
^{-}\eta \rightarrow \pi ^{-}\eta }\left( s_{23}\right) \\
&\quad + (C_{1}-C_{2}) G_{K^{0}K^{-}}\left( s_{23}\right) T_{K^{0}K^{-}\rightarrow \pi
^{-}\eta }\left( s_{23}\right) \\
&\quad + \frac{2}{\sqrt{6}} (2C_{1}+\beta C_{1}+\beta C_{2}) G_{\pi ^{+}\eta }\left( s_{13}\right) T_{\pi^{+}\eta \rightarrow \pi ^{+}\eta }\left( s_{13}\right) \\
&\quad +(C_{1}-C_{2}) G_{K^{+}\bar{K}^{0}}\left( s_{13}\right) T_{K^{+}\bar{K}^{0}\rightarrow \pi ^{+}\eta}\left( s_{13}\right),   \\
\end{aligned}
\end{equation}
where $s_{ij}$ is the energy of the two-body system in the center-of-mass frame, and the lower indices $i, j = 1, 2, 3$ represent the final states $K^+ / \pi^+(1)$, $K^- / \pi^-(2)$ and $\eta(3)$, respectively. 
Besides, the $G_{pp^{\prime}}$ is loop function of the two mesons propagator, which is logarithmically divergent. 
There are two methods to solve this problem, the three-momentum cut-off method~\cite{Oller:1997ti,Oset:1997it,Montana:2022inz,Alvarez-Ruso:2010rqm} and the dimensional regularization method~\cite{Oller:2000fj,Oller:1998zr,Gamermann:2006nm,Guo:2016zep}. 
Of course, the choice of these two regularization scheme does not affect our final conclusions. 
In this work, we adopt the three-momentum cut-off method, and the function $G_{pp^{\prime}}$ can be re-expressed as~\cite{Oset:2001cn,Guo:2005wp}, written as
\begin{equation}\label{eq14}
\begin{aligned}
G(s) &= \frac{1}{16 \pi^2 s}\left\{\sigma\left(\arctan \frac{s+\Delta}{\sigma \lambda_1}+\arctan \frac{s-\Delta}{\sigma \lambda_2}\right)\right. \\
&\quad \left.-\left[(s+\Delta) \ln \frac{\left(1+\lambda_1\right) q_{\text {max }}}{m_1}+(s-\Delta) \ln \frac{\left(1+\lambda_2\right) q_{\max }}{m_2}\right]\right\}, 
\end{aligned}
\end{equation}
where $\sigma=\left[-\left(s-\left(m_1+m_2\right)^2\right)\left(s-\left(m_1-m_2\right)^2\right)\right]^{1 / 2}$, $\Delta=m_1^2-m_2^2$, and $\lambda_i= \sqrt{1 +m_i^2 / q_{\text {max }}^2}$ $(i=1,2)$. 
$q_{max}$ is the cut-off momentum, and we take $q_{max} = 600$ MeV as used in Refs.~\cite{Dias:2016gou,Liang:2014tia,Ahmed:2021oft}. 
Furthermore, the $T_{pp \rightarrow pp}$ is the two-body scattering amplitude, which can be obtained by solving the coupled channel Bethe-Salpeter equation of the ChUA~\cite{Oller:1998zr,Oset:1997it}, given by
\begin{equation}
T = [1-VG]^{-1}V, \label{eq:BS}
\end{equation}
where $V$ is the potential matrix, constructed by the scattering potentials for every coupled channel. 
For the $I=0$ sector, there are five coupled channels: $\pi^{+} \pi^{-}(1)$, $\pi^{0} \pi^{0}(2)$, $K^{+} K^{-}(3)$, $K^{0} \bar{K}^{0}(4)$, $\eta \eta(5)$, and the elements of the $5 \times 5$ symmetric $V$ matrix can be written as~\cite{Liang:2014tia,Ahmed:2020qkv,Wang:2021ews,Gamermann:2006nm},
\begin{equation}\label{eq16}
\begin{aligned} 
V_{11} &=-\frac{1}{2 f^{2}} s, \quad V_{12}=-\frac{1}{\sqrt{2} f^{2}}\left(s-m_{\pi}^{2}\right), \quad V_{13}=-\frac{1}{4 f^{2}} s, \\  
V_{14} &=-\frac{1}{4 f^{2}} s, \quad V_{15}=-\frac{1}{3 \sqrt{2} f^{2}} m_{\pi}^{2}, \quad V_{22}=-\frac{1}{2 f^{2}} m_{\pi}^{2}, \\  
V_{23} &=-\frac{1}{4 \sqrt{2} f^{2}} s, \quad V_{24}=-\frac{1}{4 \sqrt{2} f^{2}} s, \quad V_{25}=-\frac{1}{6 f^{2}} m_{\pi}^{2}, \\ 
V_{33} &=-\frac{1}{2 f^{2}} s, \quad V_{34}=-\frac{1}{4 f^{2}} s, V_{35} =-\frac{1}{12 \sqrt{2} f^{2}}\left(9 s-6 m_{\eta}^{2}-2 m_{\pi}^{2}\right), \\ 
V_{44}&=-\frac{1}{2 f^{2}} s, \quad V_{45} =-\frac{1}{12 \sqrt{2} f^{2}}\left(9 s-6 m_{\eta}^{2}-2 m_{\pi}^{2}\right), \quad V_{55} =-\frac{1}{18 f^{2}}\left(16 m_{K}^{2}-7 m_{\pi}^{2}\right).  
\end{aligned} 
\end{equation}
For the $I=1/2$ sector, there are three channels: $K^{-} \pi^{+}(1)$, $\bar{K}^{0} \pi^{0}(2)$, $\bar{K}^{0} \eta(3)$, and then, the elements of the matrix $V$ are given by~\cite{Wang:2021ews,Toledo:2020zxj,Guo:2005wp},
\begin{equation}\label{eq17}
\begin{aligned} 
V_{11}&= \frac{-1}{6 f^{2}}\left(\frac{3}{2} s-\frac{3}{2 s}\left(m_{\pi}^{2}-m_{K}^{2}\right)^{2}\right), \\ 
V_{12}&= \frac{1}{2 \sqrt{2} f^{2}}\left(\frac{3}{2} s-m_{\pi}^{2}-m_{K}^{2}-\frac{\left(m_{\pi}^{2}-m_{K}^{2}\right)^{2}}{2 s}\right), \\ 
V_{13}&= \frac{1}{2 \sqrt{6} f^{2}}\left(\frac{3}{2} s-\frac{7}{6} m_{\pi}^{2}-\frac{1}{2} m_{\eta}^{2}-\frac{1}{3} m_{K}^{2}+\frac{3}{2 s}\left(m_{\pi}^{2}-m_{K}^{2}\right)\left(m_{\eta}^{2}-m_{K}^{2}\right)\right), \\ 
V_{22}&= \frac{-1}{4 f^{2}}\left(-\frac{s}{2}+m_{\pi}^{2}+m_{K}^{2}-\frac{\left(m_{\pi}^{2}-m_{K}^{2}\right)^{2}}{2 s}\right), \\ 
V_{23}&=-\frac{1}{4 \sqrt{3} f^{2}}\left(\frac{3}{2} s-\frac{7}{6} m_{\pi}^{2}-\frac{1}{2} m_{\eta}^{2}-\frac{1}{3} m_{K}^{2} +\frac{3}{2 s}\left(m_{\pi}^{2}-m_{K}^{2}\right)\left(m_{\eta}^{2}-m_{K}^{2}\right)\right), \\ 
V_{33}&=-\frac{1}{4 f^{2}}\left(-\frac{3}{2} s-\frac{2}{3} m_{\pi}^{2}+m_{\eta}^{2}+3 m_{K}^{2}-\frac{3}{2 s}\left(m_{\eta}^{2}-m_{K}^{2}\right)^{2}\right). 
\end{aligned} 
\end{equation}
Besides, for the $I=1$ sector, three channels are coupled, $K^{+} K^{-}(3)$, $K^{0} \bar{K}^{0}$, $\pi^{0} \eta(3)$, and the elements of the $V$ can be obtained as follows~\cite{Oller:1997ti},
\begin{equation}\label{eq18}
\begin{aligned} 
V_{11} &=-\frac{1}{2 f^{2}} s, \quad V_{12}=-\frac{1}{4f^{2}}s, \quad V_{13}=-\frac{\sqrt{3}}{12f^{2}}\left(3s-\frac{8}{3}m_{K}^2 -\frac{1}{3}m_{\pi}^{2} -m_{\eta}^{2}\right), \\  
V_{22} &=-\frac{1}{2 f^{2}} s, \quad  V_{23}=\frac{\sqrt{3}}{12f^{2}}\left(3s-\frac{8}{3}m_{K}^2 -\frac{1}{3}m_{\pi}^{2} -m_{\eta}^{2}\right), \quad V_{33} =-\frac{1}{3 f^{2}}m_{\pi}^{2},  
\end{aligned} 
\end{equation}
where $f$ is the pion decay constant which is taken as 0.093 GeV~\cite{Oller:1997ti,Toledo:2020zxj}, and $m_p$ is the mass of the corresponding pseudoscalar meson. 
Meanwhile, with the isospin multiplets $K=(K^+, K^0)$, $\bar{K}=(\bar{K}^0, -K^-)$ and $\pi=(-\pi^+, \pi^0, \pi^-)$, we can obtain the other potential with the isospin relationship, which are used in Eq.~(\ref{eq12}) and (\ref{eq13}), having
\begin{equation}
\begin{aligned} 
t_{\pi ^{-}\eta \rightarrow \pi ^{-}\eta } &=  t_{\pi ^{+}\eta \rightarrow \pi ^{+}\eta } = t_{\pi ^{0}\eta \rightarrow \pi ^{0}\eta }, \\
t_{K^{0}K^{-}\rightarrow \pi ^{-}\eta } &= t_{K^{+}\bar{K}^{0}\rightarrow \pi ^{+}\eta } = \sqrt{2}t_{K^{+}K^{-}\rightarrow \pi ^{0}\eta }=-\sqrt{2}t_{K^{0}\bar{K}^{0}\rightarrow \pi ^{0}\eta }, \\
t_{\bar{K}^{0}\pi ^{-}\rightarrow K^{-}\eta } &= t_{K^{0}\pi ^{+}\rightarrow K^{+}\eta } = t_{K^{-}\pi ^{+}\rightarrow \bar{K}^{0}\eta }=-\sqrt{2}t_{\bar{K}^{0}\pi ^{0}\rightarrow \bar{K}^{0}\eta}, \\
t_{K^{-}\pi ^{0}\rightarrow K^{-}\eta } &= t_{K^{+}\pi ^{0}\rightarrow K^{+}\eta } = \frac{1}{\sqrt{2}}t_{K^{-}\pi^{+}\rightarrow \bar{K}^{0}\eta }=-t_{\bar{K}^{0}\pi ^{0}\rightarrow \bar{K}^{0}\eta }, \\
t_{K^{-}\eta \rightarrow K^{-}\eta } &= t_{K^{+}\eta \rightarrow K^{+}\eta } = t_{\bar{K}^{0}\eta \rightarrow \bar{K}^{0}\eta }. \\
\end{aligned} 
\end{equation}

\begin{figure}[htbp]
\begin{subfigure}{0.475\textwidth}
\centering
\includegraphics[width=1\linewidth]{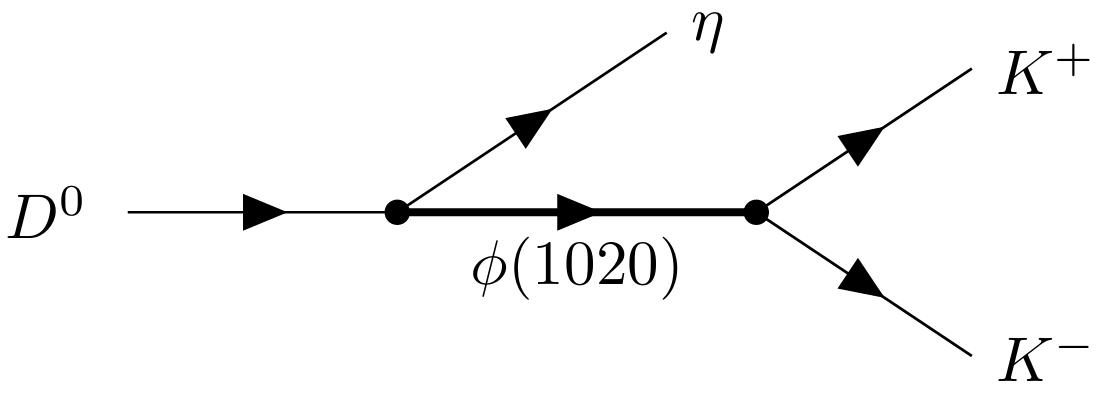} 
\caption{Mechanism via the $\phi(1020)$.}
\end{subfigure}
\begin{subfigure}{0.475\textwidth}  
\centering 
\includegraphics[width=1\linewidth]{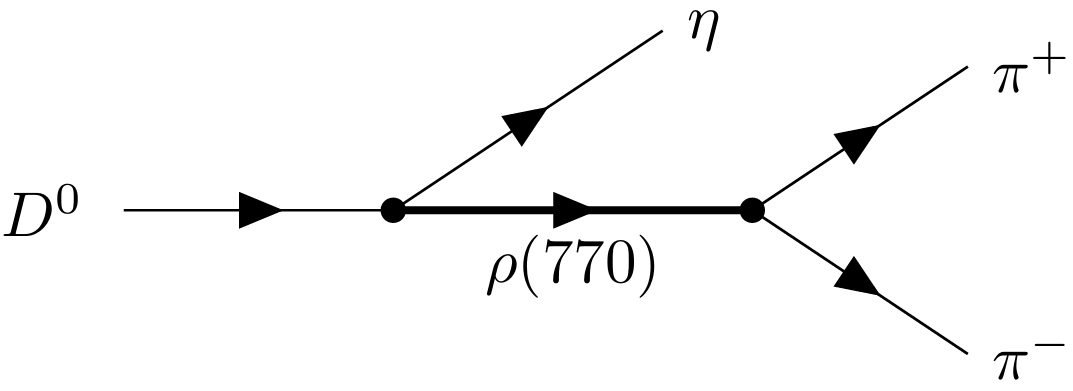} 
\caption{Mechanism via the $\rho(770)$.}
\end{subfigure}	
\captionsetup{justification=raggedright}
\caption{Mechanisms for the $D^0 \rightarrow K^+ K^- \eta$ and $\pi^+ \pi^- \eta$ processes via the intermediate vector mesons.}
\label{intermediate}
\end{figure}

Furthermore, we also consider the contribution of the $P$ wave resonances. For the $D^0 \rightarrow K^+ K^- \eta$ processes, the case of $W$-internal emission as shown in Fig.~\ref{internal}, not only can obtain the $K^+ K^-$ final states in $S$-wave through the hadronization, but also produce the $\phi(1020)$ meson in $P$-wave which decays into $K^+ K^-$. 
Similarly, for the $D^0 \rightarrow \pi^+ \pi^- \eta$ processes, the final states also can be obtained by $\rho$ intermediate state, following by $\rho \rightarrow \pi^+ \pi^-$. 
These production mechanisms are dipicted in Fig.~\ref{intermediate}, and the full relativistic amplitudes can be written as~\cite{Toledo:2020zxj,Roca:2020lyi,Liang:2023ekj},
\begin{eqnarray}
M_{\phi }\left( s_{12},s_{23}\right) = \frac{D_{\phi }e^{i\alpha _{\phi }}}{s_{12}-m_{\phi }^{2}+im_{\phi }\Gamma _{\phi }}\left( s_{23}-s_{13}\right) ,
\end{eqnarray}
\begin{eqnarray}
M_{\rho }\left( s_{12},s_{23}\right) = \frac{D_{\rho }e^{i\alpha _{\rho }}}{s_{12}-m_{\rho }^{2}+im_{\rho }\Gamma _{\rho }}\left( s_{23}-s_{13}\right) ,
\end{eqnarray}
where the $D_{\phi(1020) / \rho(770)}$ and $\alpha_{\phi(1020) / \rho(770)}$ are the normalization constants and phases, respectively, of which the values can be calculated by fitting the experimental data. 
The $\Gamma_{\phi(1020) / \rho(770)}$ and $m_{\phi(1020) / \rho(770)}$ are the total widths and masses of the corresponding intermediate states, of which the values are taken from PDG~\cite{ParticleDataGroup:2024cfk}. 

Finally, the double differential width distributions for these two three-body decays can be calculated by,
\begin{equation}
\begin{aligned} 
\frac{d^{2}\Gamma }{ds_{12}ds_{23}}=&\frac{1}{\left( 2\pi \right) ^{3}} \frac{1}{32m_{D^{0}}^{3}}  \left\vert t_{D^{0} \rightarrow \pi^{+} \pi^{-} \eta} \left( s_{12},s_{23}\right) + M_{\rho}
\right\vert ^{2}  ,
\end{aligned}  \label{integrate1}
\end{equation}
\begin{equation}
\begin{aligned} 
\frac{d^{2}\Gamma }{ds_{12}ds_{23}}=&\frac{1}{\left( 2\pi \right) ^{3}} \frac{1}{32m_{D^{0}}^{3}}  \left\vert t_{D^{0} \rightarrow K^{+} K^{-} \eta} \left( s_{12},s_{23}\right) + M_{\phi}
\right\vert ^{2}  .
\end{aligned}  \label{integrate2}
\end{equation}
Thus, the single invariant mass distributions $d\Gamma/ds_{12}$ and $d\Gamma/ds_{23}$ can be evaluated by integrating the other invariant mass invariable in Eq.~(\ref{integrate1}) and (\ref{integrate2}). 
Furthermore, the $d\Gamma/ds_{13}$ can be obtained through the constraint condition,$s_{12} + s_{13} + s_{23} = m^2_{D^0} + m^2_{K^+/\pi^+} + m^2_{K^-/\pi^-} + m^2_{\eta}$. The explicit form for the limits of integration can be taken from PDG and other literature~\cite{ParticleDataGroup:2024cfk,Liang:2023ekj,Debastiani:2016ayp}.   

Furthermore, as discussed in Refs.~\cite{Wang:2021ews,Toledo:2020zxj,Debastiani:2016ayp}, the effective energy range in the ChUA has limitations. In order to make reliable predictions up to higher energy region, we need to expand the scattering amplitudes above the energy cut $\sqrt{s_{cut}}=1.1$ GeV smoothly~\cite{Debastiani:2016ayp}, 
\begin{eqnarray}
G(s)T(s)=G(s_{cut})T(s_{cut})e^{-\alpha(\sqrt{s}-\sqrt{s_{cut}})},\quad \text { for } \sqrt{s} > \sqrt{s_{cut}},
\end{eqnarray}
where $G$ is the loop function of meson-meson propagator, which is introduced in Eq.~(\ref{eq14}). 
$T$ is the two-body scattering amplitude of the corresponding coupled channel, which is calculated by the ChUA and presented in Eq.~\eqref{eq:BS}. 
The $\alpha$ is extrapolation parameter, which is determined by fitting experimental data.

\section{Results}

In this section, we present our results obtained with the above formalism. 
Actually, we have six free parameters for each three-body process, that is, $C_1$, $C_2$, $\beta$ and $\alpha$ in $S$ wave, $D_{\phi / \rho}$ and $\alpha_{\phi / \rho}$ for the contributions of intermediate resonances from $P$ wave, of which the values can be determined by fitting the experimental data. 
As shown in Eqs.~(\ref{integrate1}) and (\ref{integrate2}), we consider the coherence of the amplitudes between the $S$ and $P$ waves.

\begin{figure}[!htbp]
\begin{subfigure}{0.52\textwidth}
\centering
\includegraphics[width=1\linewidth]{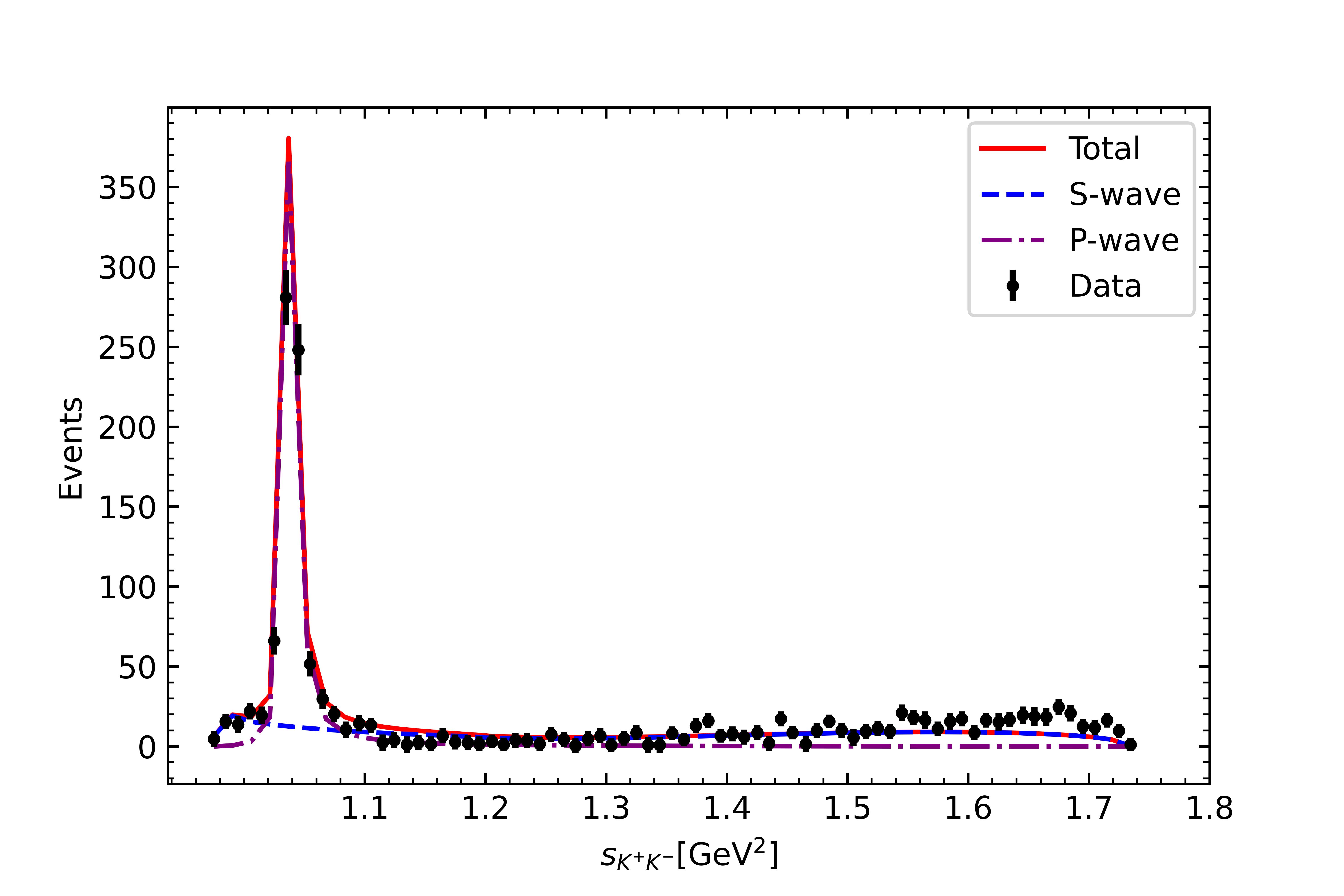} 
\caption{$K^+ K^-$ invariant mass distribution.}
\label{Contribution1}
\end{subfigure}
\begin{subfigure}{0.475\textwidth}  
\centering 
\includegraphics[width=1\linewidth]{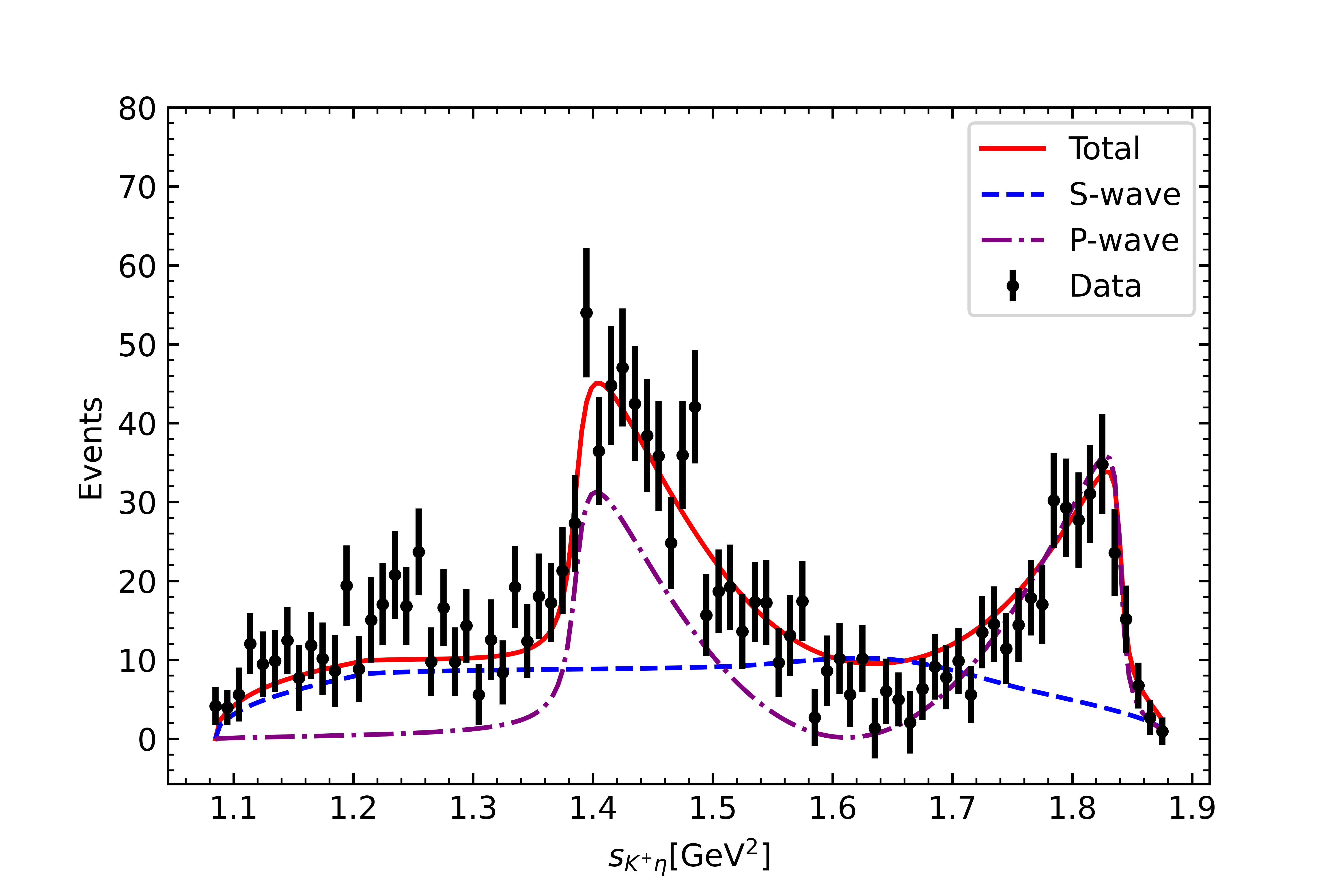} 
\caption{$K^+ \eta$ invariant mass distribution.}
\label{Contribution2}  
\end{subfigure}	
\begin{subfigure}{0.475\textwidth}  
\centering
\includegraphics[width=1\linewidth]{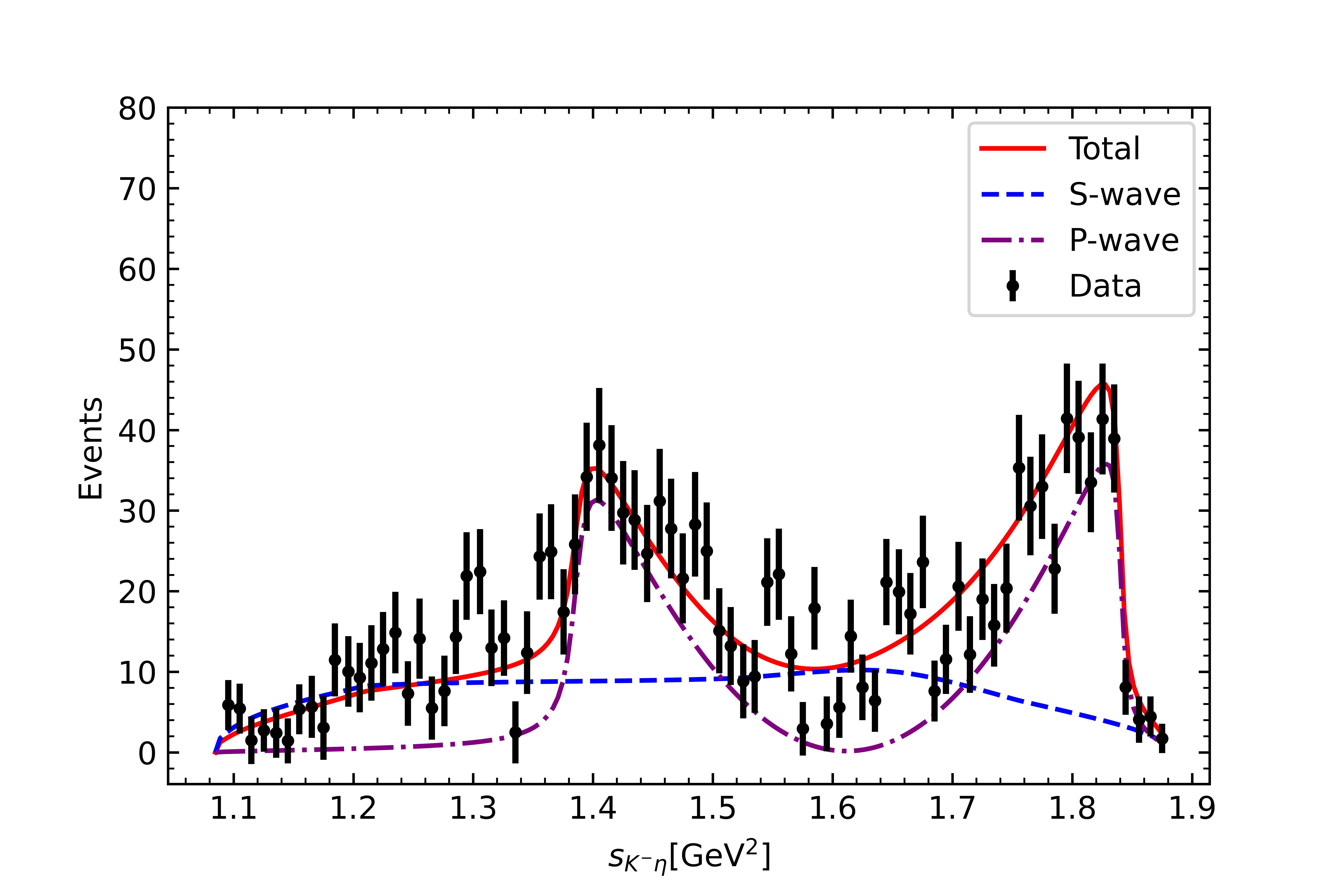} 
\caption{$K^- \eta$ invariant mass distribution.}
\label{Contribution3}
\end{subfigure}
\captionsetup{justification=raggedright}
\caption{Invariant mass distributions of the decay $D^0 \rightarrow K^+ K^- \eta$. The solid (red) line is the total contributions of the $S$ and $P$ waves, the dash (blue) line represents the $S$-wave contributions, the dash-dot (purple) line means the contributions of the $\phi(1020)$ in $P$ wave. The dot (black) points are the experimental data measured by the Belle Collaboration~\cite{Belle:2021dfa}.}
\label{fig-KKeta}
\end{figure}

For the $D^0 \rightarrow K^+ K^- \eta$ process, we perform a combined fit to the experimental data of the Belle Collaboration~\cite{Belle:2021dfa} to simultaneously describe the invariant mass distributions of $K^+ K^-$, $K^+ \eta$ and $K^- \eta$.
The fitted parameters are listed in Table.~\ref{parameters-KKeta}, and the obtained invariant mass spectra are shown in Fig.~\ref{fig-KKeta}, where one can see that our fitting results are in good agreement with the experimental data, considering the contributions from $S$ and $P$ waves. 
Additionally, with these parameters, we also can distinguish the contributions of the $S$ wave and the intermediate resonance $\phi(1020)$ from $P$ wave in the invariant mass distributions, as presented in different lines in Fig.~\ref{fig-KKeta}, see the dash (blue) and dash-dot (purple) lines, respectively. 
For the $K^+ K^-$ mass spectrum as shown in Fig.~\ref{Contribution1}, we can see a clear peak around $1.0 \text{ GeV}^2$, which corresponds to the contribution from the $\phi(1020)$, and a bit enhancement close to the $K\bar{K}$ threshold, which is the signal of the $a_0(980)$ and/or $f_0(980)$. 
For the $K^+ \eta$ and $K^- \eta$ invariant mass distributions as depicted in Figs.~\ref{Contribution2} and \ref{Contribution3}, their results are similar and show a double hump structure, around the energy regions of $1.4 \text{ GeV}^2$ and $1.8 \text{ GeV}^2$. 
One can find that the hump structure comes from the interference effect between the $S$ wave and the $\phi(1020)$ resonance of $P$ wave, where the $\phi(1020)$ are dominant.

\begin{table}[!htb]
\centering
\caption{Values of the parameters from combined fit for the data of decay $D^0 \rightarrow K^+ K^- \eta$, measured by the Belle Collaboration~\cite{Belle:2021dfa}.} \label{parameters-KKeta}
\resizebox{0.75\textwidth}{!}
{\begin{tabular}{ccccccccc}
\hline\hline
Parameters  & $C_1$ & $C_2$ & $\beta$ &  $\alpha$ & $D_{\phi(1020)}$ & $\alpha_{\phi(1020)}$  & $\chi^2/dof.$\\
\hline
Fit  &  -3448.24   & 99.07  & 0.77 & 6.28 & 123.75 & 2.89 & 1.48 \\
\hline\hline
\end{tabular}}
\end{table}

\begin{figure}[!htbp]
\begin{subfigure}{0.52\textwidth}
\centering
\includegraphics[width=1\linewidth]{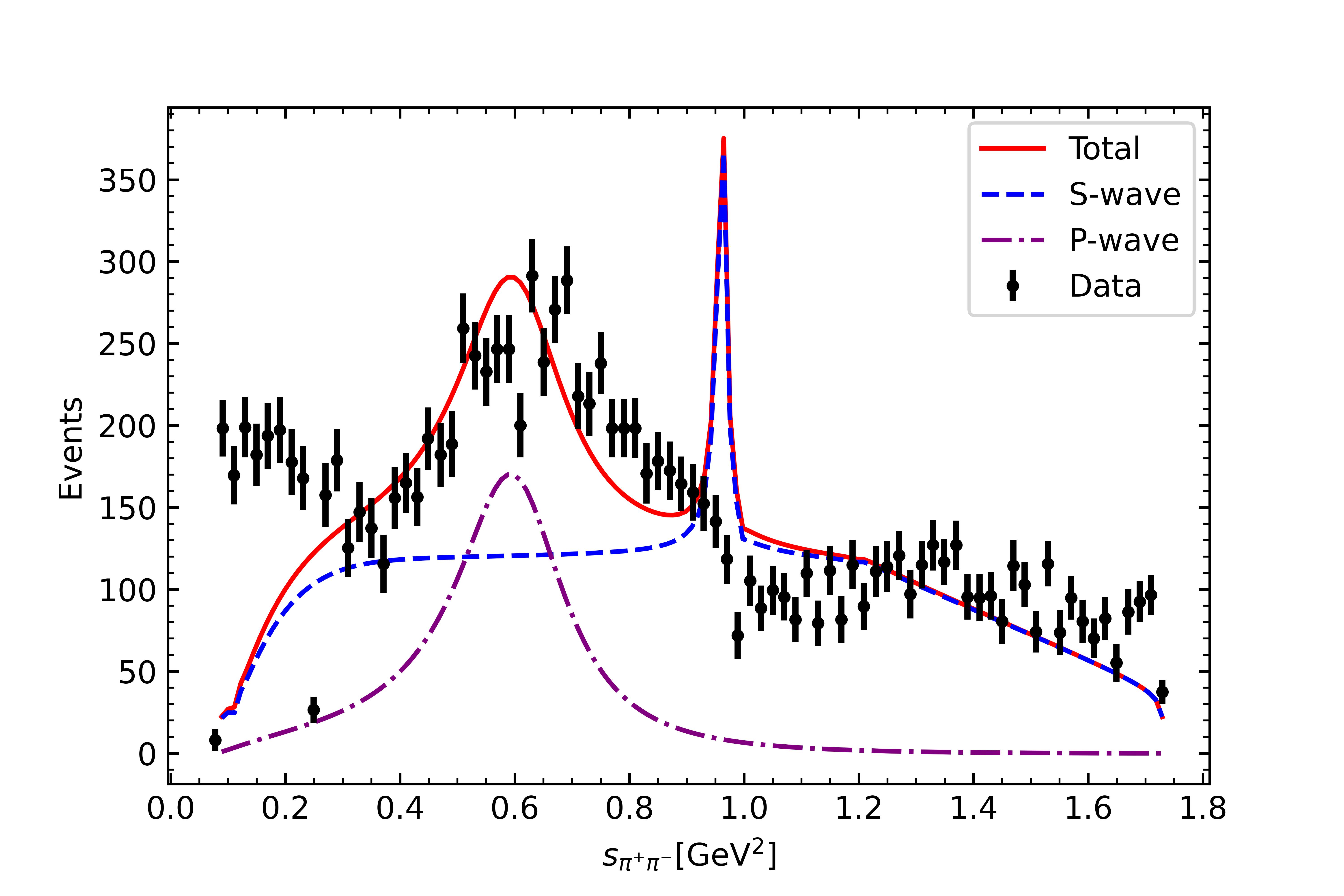} 
\caption{$\pi^+ \pi^-$ invariant mass distribution.}
\label{Contribution4}
\end{subfigure}
\begin{subfigure}{0.475\textwidth}  
\centering 
\includegraphics[width=1\linewidth]{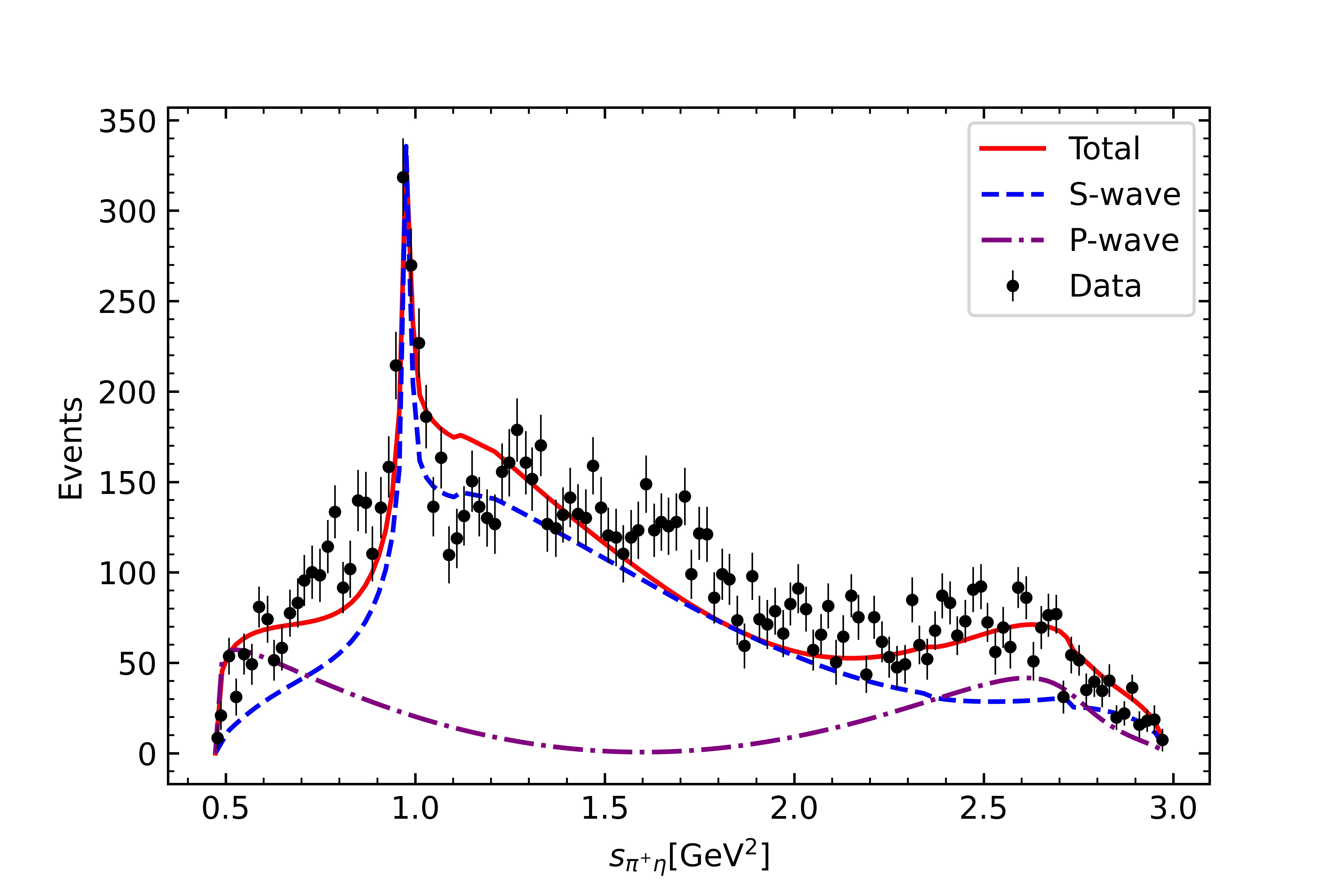} 
\caption{$\pi^+ \eta$ invariant mass distribution.}
\label{Contribution5}  
\end{subfigure}	
\begin{subfigure}{0.475\textwidth}  
\centering
\includegraphics[width=1\linewidth]{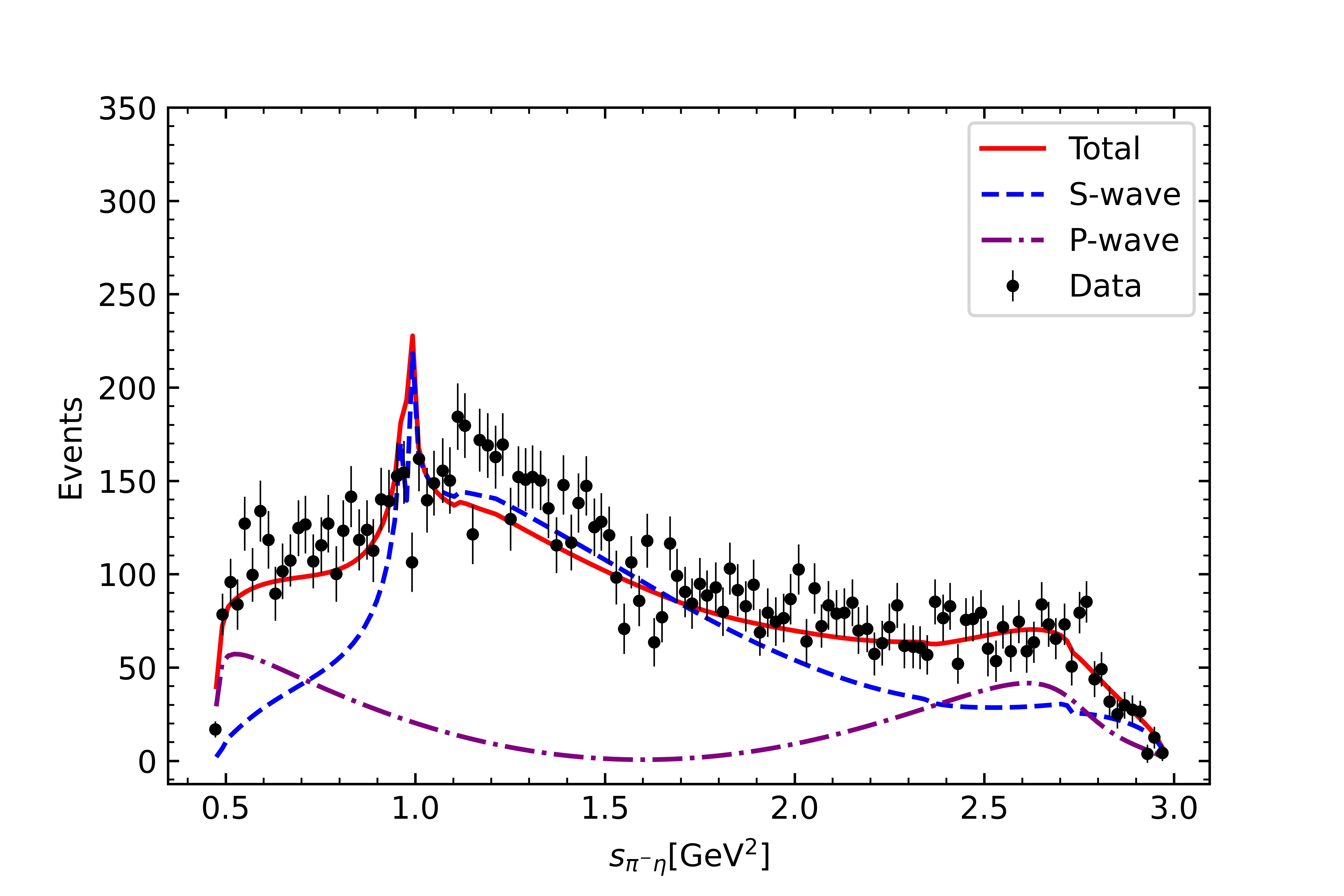} 
\caption{$\pi^- \eta$ invariant mass distribution.}
\label{Contribution6}
\end{subfigure}
\captionsetup{justification=raggedright}
\caption{Invariant mass distributions of the decay $D^0 \rightarrow \pi^+ \pi^- \eta$. The solid (red) line is the total contributions of the $S$ and $P$ waves, the dash (blue) line represents the $S$-wave contributions, the dash-dot (purple) line means the contributions of the $\rho(770)$. The dot (black) points are the experimental data measured by the Belle Collaboration~\cite{Belle:2021dfa}.}
\label{fig-pipieta}
\end{figure}

For the $D^0 \rightarrow \pi^+ \pi^- \eta$ process, we first make a combined fit with the data of the $\pi^+ \pi^-$, $\pi^+ \eta$ and $\pi^- \eta$ invariant mass distributions measured by the Belle Collaboration~\cite{Belle:2021dfa}. 
The fitted parameters are given in Table~\ref{parameters-pipieta}, and the corresponding invariant mass spectra with these parameters are shown in Fig.~\ref{fig-pipieta}. 
From Fig.~\ref{fig-pipieta}, one can find that, except for the part of data at the low energy region above the threshold in the $\pi^+ \pi^-$ distribution~\footnote{This part can not be described well even doing single fit for the $\pi^+ \pi^-$  invariant mass distributions.}, which leads to a bit large of the value $\chi^2/dof.$, the other fitted results are agree with the experimental data within the error ranges. 
In Fig.~\ref{Contribution4}, there is a distinct peak appeared in the $\pi^+ \pi^-$ spectrum around $1.0 \text{ GeV}^2$, corresponding to the $f_0(980)$ resonance, which is the contribution of $I=0$ sector encoded in Eq.~(\ref{eq13}), while, this signal is not visible in experimental results~\cite{Belle:2021dfa}. 
Note that the contributions of the state $f_0(500)$ from $S$ wave can be found around the region of $0.2\sim0.4 \text{ GeV}^2$, and the ones of the $\rho$ in $P$ wave is dominant in the region of $0.6 \text{ GeV}^2$. 
In Fig.~\ref{Contribution5}, the data of the $\pi^+ \eta$ invariant mass distribution is well fitted, and the peak around $1.0 \text{ GeV}^2$ is dominant by the contribution of the $a_0(980)$, which comes from the interaction of the $I=1$ sector. 
Note that the contribution from the $a_0(980)$ is relatively large than the one from the $\rho(770)$. 
Besides, the invariant mass distribution of the $\pi^- \eta$ is exhibited in Fig.~\ref{Contribution6}, which is similar to what we has for the one $\pi^+ \eta$. 
As one can see the feature from Figs.~\ref{Contribution5} and~\ref{Contribution6} that, with one set of the combined fit parameters, both of the $\pi^+ \eta$ and $\pi^- \eta$ mass distributions are described well, even though they are not isospin symmetry, as indicating by the BESIII Collaboration~\cite{BESIII:2024tpv} that the production of the $a_0(980)^+$ is larger than the one of the $a_0(980)^-$.

\begin{table}[!htb]
\centering
\caption{Values of the parameters from combined fit for the data of decay $D^0 \rightarrow \pi^+ \pi^- \eta$, measured by the Belle Collaboration~\cite{Belle:2021dfa}.} \label{parameters-pipieta}
\resizebox{0.75\textwidth}{!}
{\begin{tabular}{ccccccccc}
\hline\hline
Parameters  & $C_1$ & $C_2$ & $\beta$ &  $\alpha$ & $D_{\rho(770)}$ & $\alpha_{\rho(770)}$  & $\chi^2/dof.$\\
\hline
Fit  &  419.79   & 2196.24  & 1.00 & 0.38 & -170.97 & 2.20 & 5.23 \\
\hline\hline
\end{tabular}}
\end{table}

\begin{figure}[!htbp]
\begin{subfigure}{0.52\textwidth}
\centering
\includegraphics[width=1\linewidth]{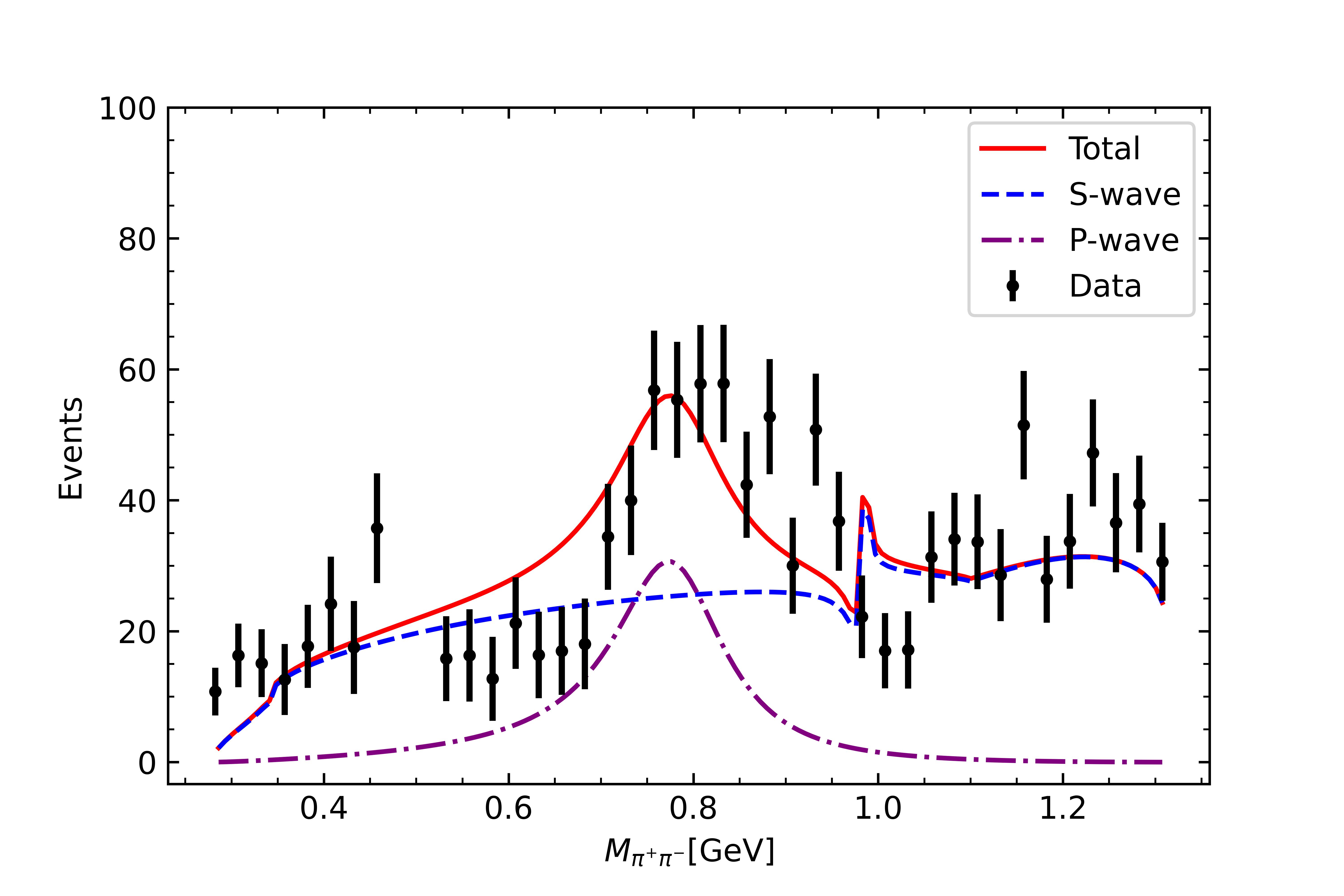} 
\caption{Invariant mass distribution of $\pi^+ \pi^-$.}
\label{Contribution7}
\end{subfigure}
\begin{subfigure}{0.475\textwidth}  
\centering 
\includegraphics[width=1\linewidth]{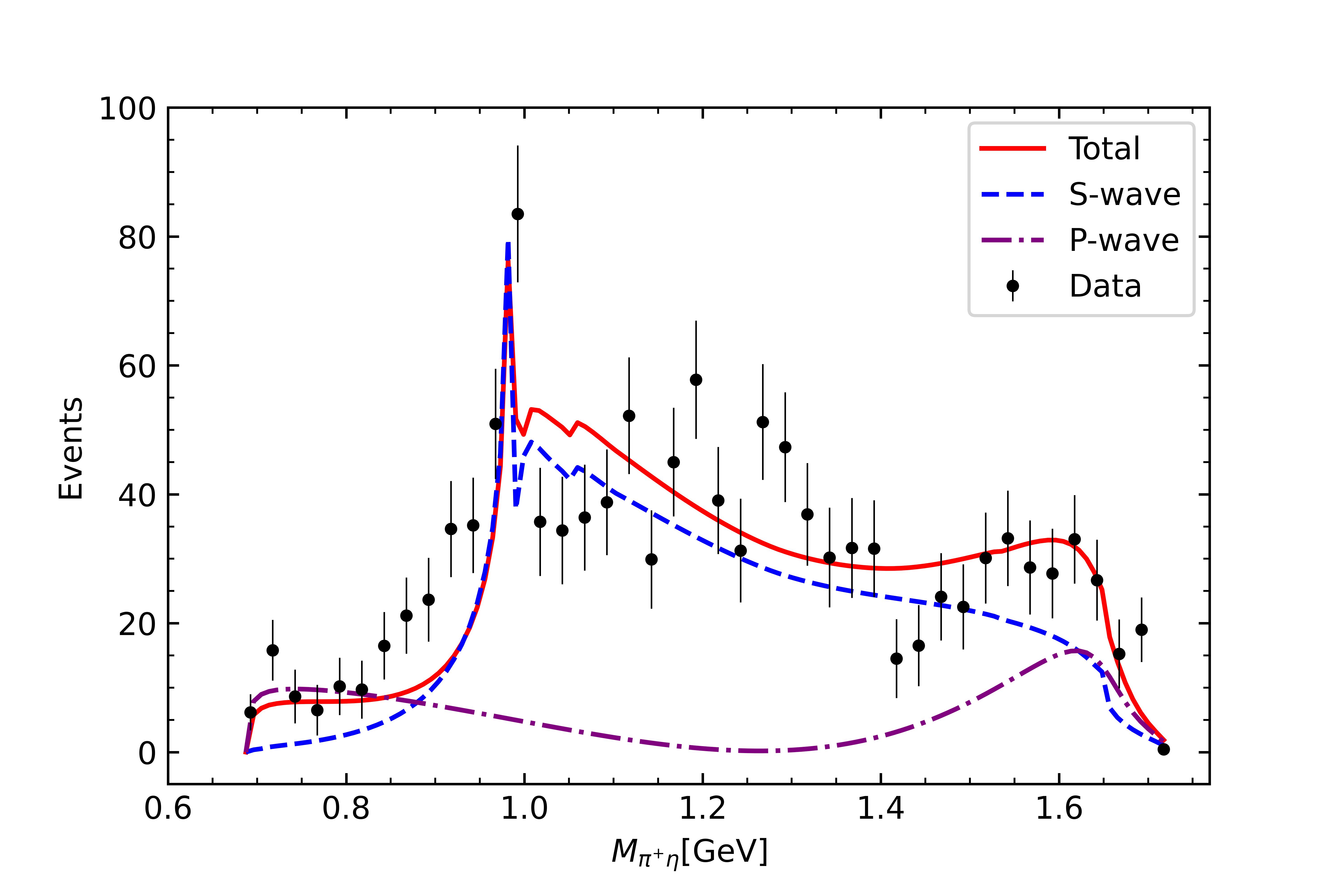} 
\caption{Invariant mass distribution of $\pi^+ \eta$.}
\label{Contribution8}  
\end{subfigure}	
\begin{subfigure}{0.475\textwidth}  
\centering
\includegraphics[width=1\linewidth]{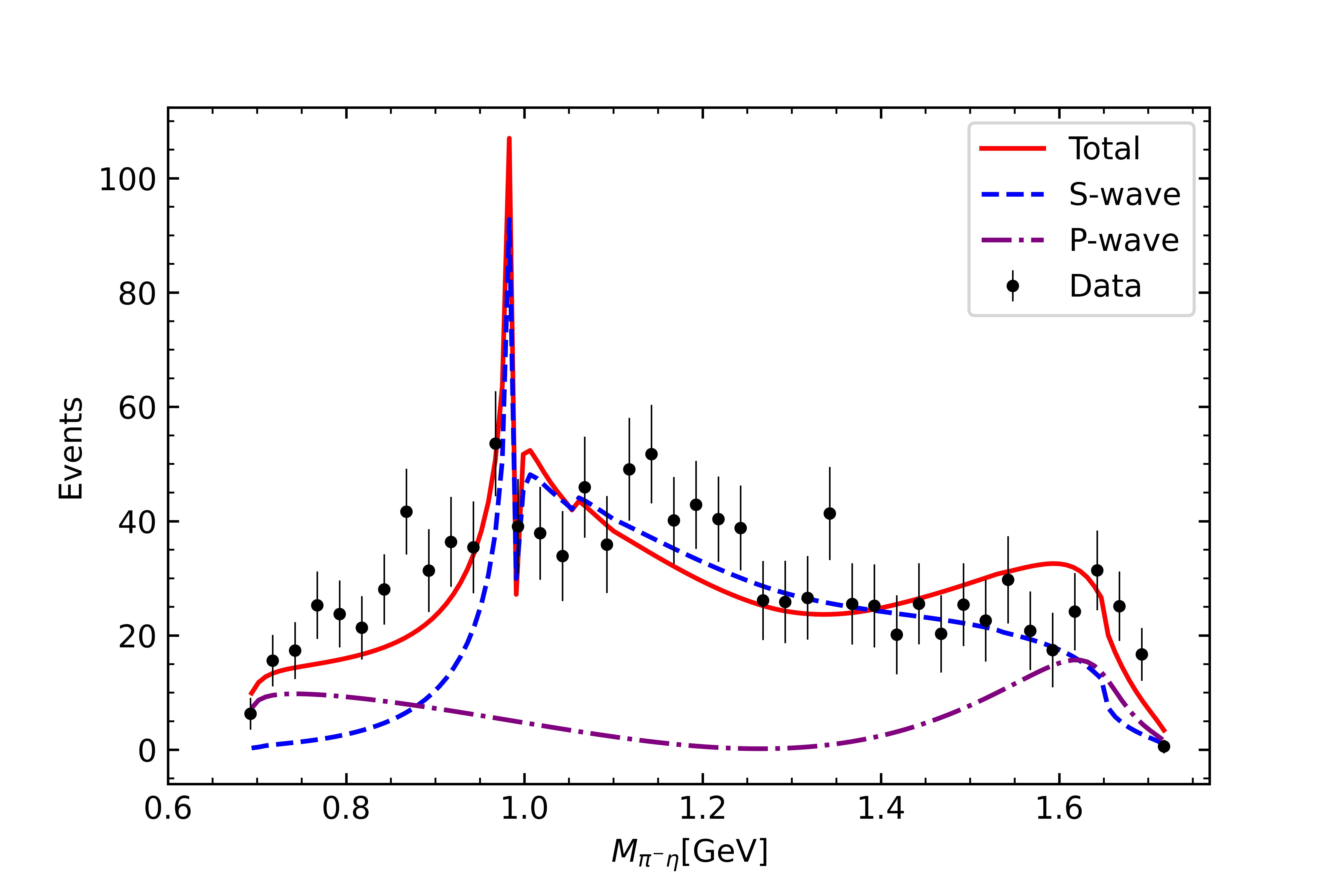} 
\caption{Invariant mass distribution of $\pi^- \eta$.}
\label{Contribution9}
\end{subfigure}
\captionsetup{justification=raggedright}
\caption{Combined fit for the invariant mass distributions of the decay $D^0 \rightarrow \pi^+ \pi^- \eta$. The dot (black) points are the experimental data measured by the BESIII Collaboration~\cite{BESIII:2024tpv}, Others are the same as Fig~\ref{fig-pipieta}.}
\label{fig-pipieta-BESIII}
\end{figure}

Since recently the BESIII Collaboration also investigated the $D^0 \rightarrow \pi^+ \pi^- \eta$ decay and reported corresponding invariant mass distributions~\cite{BESIII:2024tpv}, we make a furrther combined fit with their measured data. 
The fitted parameters are listed in Table~\ref{parameters-pipieta-BESIII}, and the reproduced mass spectra are presented in Fig.~\ref{fig-pipieta-BESIII}, where our fit results show great agreement with experiments within the error range. 
One can see that the fitted results for these three distributions are similar with the results in Fig.~\ref{fig-pipieta}. 
In Fig.~\ref{Contribution7}, the contributions of the state $f_0(500)$ in $S$ wave and the $\rho$ in $P$ wave are also found in the corresponding energy regions. 
But, now the signal of the $f_0(980)$ resonance in the $\pi^+ \pi^-$ spectrum around $1.0 \text{ GeV}$, see Fig.~\ref{Contribution7}, is not so strong compared to the one in Fig.~\ref{Contribution4}, and thus, the value $\chi^2/dof.$ in Table~\ref{parameters-pipieta-BESIII} becomes one half smaller than the one in Table~\ref{parameters-pipieta}. 
Indeed, the theoretical signal for the $f_0(980)$ resonance is not so visible compared with the one of the state $f_0(500)$, where these signals were claimed as the contributions of $\pi^+\pi^-$ $S$-wave scattering in the experiment~\cite{BESIII:2024tpv}. 
Of course, the feature is still kept, that the isospin antisymmetry data for the production of the $a_0(980)^+$ and $a_0(980)^-$ is also fitted well with one set of the combined fit parameters, as shown in Figs.~\ref{Contribution8} and~\ref{Contribution9}.

\begin{table}[!htb]
\centering
\caption{Values of the parameters from combined fit for the data of decay $D^0 \rightarrow \pi^+ \pi^- \eta$, measured by the BESIII Collaboration~\cite{BESIII:2024tpv}.} \label{parameters-pipieta-BESIII}
\resizebox{0.75\textwidth}{!}
{\begin{tabular}{ccccccccc}
\hline\hline
Parameters  & $C_1$ & $C_2$ & $\beta$ &  $\alpha$ & $D_{\rho(770)}$ & $\alpha_{\rho(770)}$  & $\chi^2/dof.$\\
\hline
Fit  &  282.74   & -516.33  & 1.00 & 2.47 & 58.41 & 4.36 & 2.52 \\
\hline\hline
\end{tabular}}
\end{table}

\section{Summary}\label{Sec4}

Based on the experimental results for the $D^0 \rightarrow K^+ K^- \eta$ and $\pi^+ \pi^- \eta$ three-body weak decays measured by the Belle and BESIII Collaborations, we adopt the chiral unitary approach and final state interaction formalism to investigate these two processes. 
Taking into account the contribution of the $S$-wave interactions between pseudoscalar and pseudoscalar mesons, the $a_0(980)$ resonance is dynamically generated in the final state interactions. 
Besides, we also consider the contributions of intermediate resonances in $P$ wave, that is, the $\phi$ meson for the $D^0 \rightarrow K^+ K^- \eta$ and the $\rho$ meson for the $D^0 \rightarrow \pi^+ \pi^- \eta$. 
Considering the coherence effect between the amplitudes of the $S$ and $P$ wave, we do a combined fit of the experimental data to determine the free parameters of the formalism. 
Most of results are in good agreement with the experimental data, indicating the contributions of the $a_0$(980) in $S$ wave and the $\phi$ or $\rho$ in $P$ wave are great significance for these two decays. 
Note that, the antisymmetry data of the $\pi^+ \eta$ and $\pi^- \eta$ mass distributions are described well in our combined fit.

\section*{Acknowledgements}

We thank Long-Ke Li for the useful discussions. 
This work is supported by the Fundamental Research Funds for the Central Universities of Central South University under Grants No. 1053320214315, 2022ZZTS0169, and the Postgraduate Scientific Research Innovation Project of Hunan Province under No. CX20220255, 
and partly by the Natural Science Foundation of Hunan province under Grant No. 2023JJ30647, 
the Natural Science Foundation of Guangxi province under Grant No. 2023JJA110076, 
and the National Natural Science Foundation of China under Grant Nos. 12365019, 12205384, 12275076, 11905258, 12335002.

\end{document}